\documentclass[11pt,a4paper]{article}
\pdfoutput=1
\usepackage{jheppub}
\usepackage{gensymb}
\usepackage{placeins}
\usepackage{comment}
\usepackage{subfigure}
\usepackage{amssymb,amsmath}
\usepackage{graphicx}
\usepackage{color}
\usepackage{cancel}
\usepackage[colorlinks=true
,urlcolor=blue
,citecolor=blue
,linkcolor=blue
,linktocpage=true
,pdfproducer=medialab
]{hyperref}

\usepackage[numbers]{natbib}
\usepackage{notoccite}
\makeatletter \renewcommand{\@dotsep}{10000} \makeatother
\def\be{\begin{equation}}
\def\ee{\end{equation}}
\def\bea{\begin{eqnarray}}
\def\eea{\end{eqnarray}}
\def\bi{\begin{itemize}}
\def\ei{\end{itemize}}

\usepackage[nodisplayskipstretch]{setspace}
\newcommand{\mgut}{M_{{\rm GUT}}}

\DeclareUnicodeCharacter{2212}{-}
\DeclareUnicodeCharacter{202F}{\,}
\begin{document}

\begin{titlepage}
\pagestyle{empty}

\vspace*{0.2in}
\begin{center}
{\Large \bf Muon g-2 and lepton flavor violation in supersymmetric GUTs}

\vspace{1 true cm}

  Mario E. G\'{o}mez$^{a,}$\footnote{Email: mario.gomez@dfa.uhu.es}, Smaragda Lola$^{b,}$\footnote{E-mail: magdalola@upatras.gr}, Qaisar Shafi$^{c,}$\footnote{E-mail: qshafi@udel.edu} and
Cem Salih $\ddot{\rm U}$n$^{a,d,}\hspace{0.05cm}$\footnote{E-mail: cemsalihun@uludag.edu.tr}

\vspace{0.5cm}

{\it
$^a$Dpt. de Ciencias Integradas y Centro de Estudios Avanzados en F\'{i}sica Matem\'aticas y Computaci\'on, Campus del Carmen, Universidad de Huelva, Huelva 21071, Spain \\
$^b$Department of Physics, University of Patras, 26500 Patras, Greece\\
$^c$Department of Physics and Astronomy,
University of Delaware, Newark, DE 19716, USA \\
$^d$Department of Physics, Bursa Uluda\~{g} University, TR16059 Bursa, Turkey
}

\end{center}

\vspace{0.5cm}
\begin{abstract}

We present a class of supersymmetric (SUSY) GUT models that can explain the apparent discrepancy between the SM predictions and experimental values of 
muon $g-2$ while providing testable signals for lepton flavor violation in charged lepton decays. Moreover, these models predict LSP neutralino abundance that is compatible with the Planck dark matter bounds. We find that scenarios in the framework of  
$SU(4)_c\times SU(2)_L\times SU(2)_R$ unification, with additional symmetries to explain fermion masses and neutrino oscillations,  provide interesting benchmarks for the search of SUSY by correlating a possible manifestation of it in dark matter, rare lepton decays and LHC signals.

% can explain the discrepancy between SM predictions and experimental values, while predicting a DM candidate in agreement with WMAP observations. This can be achieved in models where Supersymmetry is broken through Supergravity in such a way that soft masses are generated by TeV gaugino masses and light gravitinos. If, in addition,  these models are extended to provide an explanation for fermion masses and/or neutrino oscillations, contributions to charged lepton flavor violation (LFV) appear. We found scenarios that can provide a natural explanation to the $g-2$ problem without violating the current bounds on LFV of charged leptons, even if the soft scalar masses are flavor-dependent at the GUT scale. In this work, we study such scenarios in the framework of  
% $SU(4)_c\times SU(2)_L\times SU(2)_R$ unification, with additional symmetries to explain fermion masses. 

\end{abstract}
\end{titlepage}

%\tableofcontents
%\noindent \hrulefill

%\baselineskip 36pt
\section{Introduction}
\label{sec:intro}

In addition to providing a solution to the hierarchy problem by shielding the Higgs boson mass from quadratically divergent radiative corrections, supersymmetry (SUSY) is a well motivated extension of the Standard Model (SM) since it also offers a plausible dark matter (DM) candidate. In addition, the supersymmetric partners of the SM particles provide new contributions to the muon anomalous magnetic moment (muon $g-2$) to better fit with the current experimental measurements
\cite{Muong-2:2023cdq,Muong-2:2024hpx,Muong-2:2021ojo, Muong-2:2006rrc}. Finally, a successful unification of the SM gauge couplings in its minimal supersymmetric extension (MSSM) motivates one to build SUSY grand unified theories (GUTs) that are spontaneously broken at the grand unification scale ($M_{{\rm GUT}}\simeq 2\times 10^{16}$ GeV). A solution to the muon $g-2$ problem together with a compelling explanation of the Higgs boson mass and plausible DM candidates can lead to mass relations among the SUSY particles which can be tested in the current experiments. For instance, while the muon $g-2$ favors a relatively light SUSY mass spectrum, the observed Higgs boson mass of 125 GeV \cite{ATLAS:2012yve,CMS:2013btf} requires a fairly heavy set of SUSY particles. Some GUT relations among the soft gaugino masses can alleviate this challenge by predicting heavy gluinos, which enhance the squark masses through the renormalization group equations (RGEs), while the sleptons remain relatively light \cite{Chakraborti:2021dli,Baer:2021aax,Aboubrahim:2021xfi,Wang:2021bcx,Han:2020exx,Altin:2017sxx,Li:2021pnt,Ellis:2021zmg,Athron:2021iuf,Chakraborti:2021bmv,Endo:2021zal,Iwamoto:2021aaf,Baum:2021qzx,Frank:2021nkq,Heinemeyer:2021opc,Akula:2013ioa, Gomez:2022qrb, Ellis:2024ijt}. These schemes can be compatible with flavor-mixing scalar soft terms at the GUT scale, where the scalar masses are mostly induced by flavor-independent large gaugino masses through the RGEs \cite{Gomez:2010ga}. In these cases, flavor violation is suppressed in the quark sector, while the lepton sector can still yield interesting phenomenology, especially since it is favored by the muon $g-2$ solution.

Nevertheless, since MSSM is built on the same gauge symmetry as SM, the SM implications for flavor mixing in the charged leptons, neutrino masses and mixing remain intact in the MSSM framework. Addressing it may require additional fields and symmetries, which induce lepton flavor violation (LFV) processes. The bounds on such processes \cite{Borzumati:1986qx, Barbieri:1995tw, Goldberg:1996vd} provide a remarkable complementarity to direct SUSY searches at the Large Hadron Collider (LHC). For instance, if the MSSM is extended via a seesaw mechanism to accommodate neutrino masses and mixings, it can also lead to a misalignment between the leptons and sleptons at the loop-level and induce LFV processes in the charged lepton sector \cite{Hammad:2016bng,Abdallah:2011ew,Khalil:2009tm,Arganda:2015naa}. In addition, if the MSSM is supplemented with flavor symmetries above the GUT scale, the flavor dependence of the K\"ahler potential may enhance such a misalignment~\cite{Gomez:1998wj,King:2004tx,Olive:2008vv,Ellis:2015dra,Ellis:2016qra,Das:2016czs}.

In this work, we consider a class of SUSY GUTs based on $SU(4)_{c}\times SU(2)_{L}\times SU(2)_{R}$ gauge group (hereafter, abreviated as $4-2-2$) \cite{Pati:1974yy}, which is broken to the MSSM gauge group together with the Left-Right (LR) symmetry as described in \cite{Gomez:2022qrb}. We also assume a $U(1)$ flavor symmetry above the GUT scale. If the SUSY breaking in such models is primarily mediated by supergravity, these assumptions together lead to flavor mixing in the sfermion mass matrices and yield LFV processes \cite{Gomez:2010ga}. The implications for these processes can be strongly motivated if one explores the interplay between them and the new muon $g-2$ contributions, since the particles enhancing muon $g-2$ also contribute to the LFV processes. The solutions yielding consistent LFV implications together with the desired contributions to muon $g-2$ can be further constrained by requiring the lightest supersymmetric particle (LSP) to be a suitable DM candidate in agreement with the cosmological constraints.

The rest of the paper is organized as follows: We will review in Section \ref{sec:g2andLFV} the muon $g-2$ contributions and LFV processes with the relevant particles and parameters. We summarize the experimental constraints and selection rule for the solutions in our scans in Section \ref{sec:scan}. In Section \ref{sec:results} we present our results for LFV implications and the impact on the parameter space, muon $g-2$ and SUSY mass spectrum and some prospects for tests at the LHC probe in several subsections. After exemplifying our findings with two tables listing the benchmark points, we summarize our results in Section \ref{sec:conc}.

\section{Supersymmetric contribution to muon g-2 and LFV.}
\label{sec:g2andLFV}
\begin{figure}[htb!]
\centering
\includegraphics[scale=1.7]{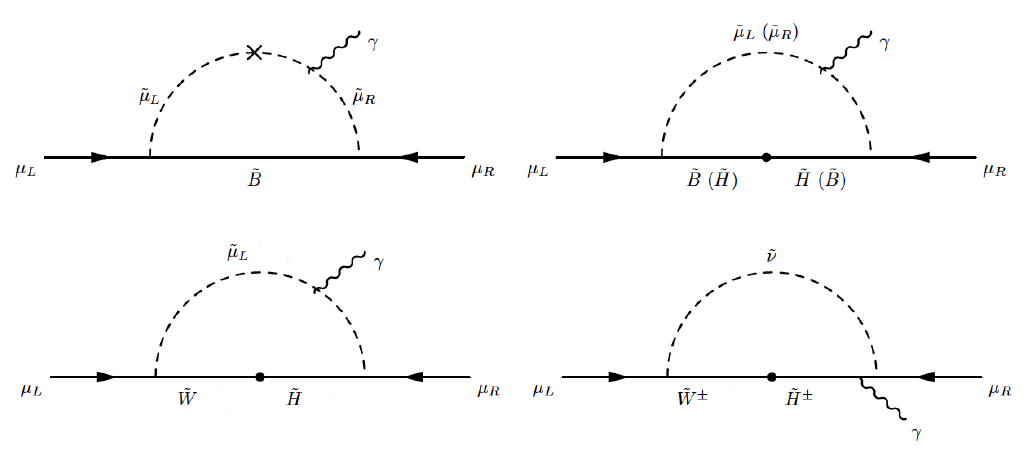}
\caption{\it{The leading contributions to  muon $g-2$ involving neutralino and chargino loops. The cross in the top-left diagram denotes the chirality flip between the left- and right-handed smuons, while the dots in the other diagrams represent the mixing between different neutralino species. In the top-right diagram, there is another loop which is formed by the particles given in the parentheses.}}
\label{fig:SNloops}
\end{figure}

\begin{figure}
\begin{center}
\includegraphics*[scale=.4]{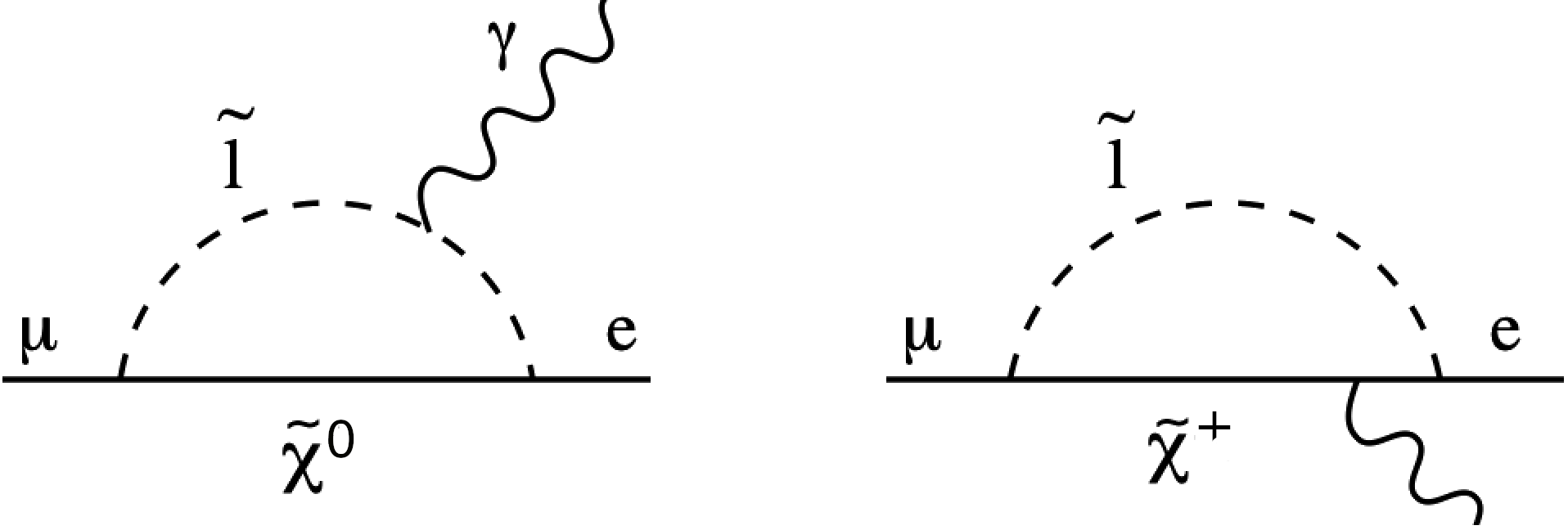}
%\vspace{-0.5 true cm}
\caption{\it One-loop diagrams for the decay $\mu\rightarrow e\gamma$: $\tilde l$ represents a charged slepton (left) or
sneutrino (right), and $\tilde\chi ^{0}$, $\tilde\chi ^{+}$ represent  neutralinos and charginos respectively.}
\label{fig:muegamma}
\end{center}
\end{figure}

The SUSY contributions to muon $g-2$ can arise at one-loop from the electroweak gauginos, Higgsinos and sleptons through the diagrams shown in Fig. \ref{fig:SNloops} \cite{Martin:2001st,Giudice:2012pf,Moroi:1995yh}, and these contributions potentially fill the gap between the latest measurements \cite{Muong-2:2024hpx} and the muon $g-2$ predctions SM \cite{Aoyama:2020ynm}. However, the current experimental constraints, especially from the Higgs boson mass, can exclude the required relatively  large contributions in the simplest SUSY models such as CMSSM. In \cite{Gomez:2022qrb} we have shown that MSSM models derived from $4-2-2$ scheme described above can explain muon $g-2$. Here, we will discuss in addition that when leptons and sleptons mass matrices are not aligned in the same superfield basis, Lepton Flavor Violating (LFV) processes can occur through very similar topologies as shown in Fig. \ref{fig:muegamma}. In this section, we  consider the muon $g-2$ and LFV processes and emphasize the interplay between them. 

\subsection{ A ${4-2-2}$ SUSY-GUT framework.}
In this subsection we review some  aspects of the $4-2-2$ symmetry relevant for extracting the MSSM 
framework that we will use in our computations. The  $4-2-2$  is one of the symmetries containing the SM \cite{Pati:1974yy,Lazarides:1980tg,Kibble:1982ae} that can be preserved in different breaking schemes of $SO(10)$ \cite{Babu:1992ia,Anderson:1993fe,Drees:1986vd,Kawamura:1993uf,Kolda:1995iw,Miller:2012vn,Babu:2005ui,Milagre:2024wcg}, for instance, through the vacuum expectation values (VEVs) of the Higgs fields residing either in $54_{H}$ or $210_{H}$ Higgs multiplets. The breaking through the VEV in $54_{H}$ preserves the so called $C-$parity \cite{Kibble:1982dd,Lazarides:1985my}, which transforms the left-handed and right-handed fields into each other. However, if one implements the $SO(10)$ breaking through the $210_{H}$ VEV, the $C-$parity is broken \cite{Babu:2016bmy}. In this work we will adopt the latter scheme, assuming left and right fields belong to disconnected $4-2-2$ representations. 

The breaking of $4-2-2$ to the MSSM can be driven at the GUT scale by the Higgs fields\cite{King:1997ia}: 
\begin{eqnarray}
\bar{H}&=&(\bar{4}, 1, \bar{2})=(u^c_H, d^c_H,\nu^c_H,e^c_H) \nonumber \\
H&=&(4, 1,2)=(\bar{u}^c_H, \bar{d}^c_H,\bar{\nu}^c_H,\bar{e}^c_H)
\end{eqnarray}
through the vevs of their neutral components $\left<\nu^c_H\right>=\left<\bar{\nu}^c_H\right>\sim M_{GUT}$,   
leaving intact the hypercharge generator $Y$, where

\begin{equation}
Y = \sqrt{\dfrac{3}{5}} I_{3R} + \sqrt{\dfrac{2}{5}}(B-L)~,
\end{equation}
\noindent
here $I_{3R}$ and $B-L$ represent the diagonal generators of $SU(2)_{R}$ and $SU(4)_{c}$, respectively. Taking into account supersymmetry breaking yields the following relation among  the three soft-susy breaking (SSB) gaugino masses at the GUT scale:

\begin{equation}
M_{1}=\dfrac{3}{5}M_{2R} + \dfrac{2}{5}M_{4}~,
\label{eq:PSgauginos}
\end{equation}
where $M_{2R}$ and $M_{4}$ denote the gaugino mass terms for $SU(2)_{R}$ and $SU(4)_{c}$, respectively, and $M_{3}=M_{4}$ at $\mgut$. If the LR breaking  in the gaugino sector is parametrised as $M_{2R}=y_{LR}M_{2L}$, where $M_{2L}$ is the SSB mass  of $SU(2)_{L}$ gaugino, the SSB gaugino masses become independent. This is crucial in order to find 
a supersymmetric solution to the muon $g-2$, since typically models with universal gaugino masses add a small contribution to it. This is a consequence of the large SUSY spectrum resulting from the gluino mass bound. In contrast, non-universal gaugino masses contribute selectively to a SUSY spectrum where large squark masses coexists with lighter sleptons that can explain the muon $g-2$ anomaly. 

\begin{comment}
    
{\it This non-universality in gaugino masses helps to 
remove the tension between the supersymmetric muon $g-2$ contributions and the fairly severe gluino mass bound that would . Because, the gluino contribution to squarks results into a large mass splitting of those respect the sfermions. 
}
\end{comment}

The squarks and leptons reside in the representations: 
\begin{eqnarray}
    F_i=(4,2,1)_i&=&\left(\begin{array}{cccc} u_r& u_g &u_b& \nu_l\\
    d_r& d_g &d_b& e_l \end{array}
    \right)_i,\nonumber\\
 \bar{F}_i=(\bar{4},2,1)_i&=&\left(\begin{array}{cccc} u^c_r& u^c_g &u^c_b& \nu^c_l\\
    d^c_r& d^c_g &d^c_b&e^c_l\end{array}
    \right)_i,
\end{eqnarray}
\noindent
where $i=1,2,3$ is the family index and $r,g,b,l$ denote the 4 colors of $SU(4)_C$. 

\begin{comment}
The MSSM electroweak Higgs doublets are contained in the representation: 
\begin{equation}
    h=(1,2,\bar{2})=\left( \begin{array}{cc} h_d^0 &h_u^+ \\
   h_d^- &h_u^0 \end{array}
    \right),
\end{equation}
At the electroweak scale, the neutral components develop the vev's $v_u=\left<h^0_u\right>$ and $v_d=\left<h^0_c\right>$, defining $tan\beta=v_d/v_u$. 
\end{comment}

The broken LR symmetry, in general, implies non-universal masses for the soft left- and right-handed matter scalar masses which can be quantified as $m_{R}\equiv x_{LR}\cdot m_{L}$, where $m_{R}$ ($m_{L}$) denotes the SSB mass of the right-handed (left-handed)  fields.  In this work, we assume that the $4-2-2$ symmetry is broken to the MSSM gauge group near the GUT scale, and thus we require the solutions to maintain approximately unification of the MSSM gauge couplings at $\mgut$. The $4-2-2$ particle spectrum contains right-handed neutrinos, which play a role in explaining  neutrino oscillations established by the current experiments \cite{Super-Kamiokande:2010orq}. However, an extenision of the MSSM with tiny neutrino masses by a {\em see-saw} mechanism requires either very heavy right-handed neutrinos or negligible couplings between the MSSM particles and the right-handed neutrinos ($Y_{\nu} \lesssim 10^{-7}$) \cite{Coriano:2014wxa,Khalil:2010zza,Abbas:2007ag}, and so the right-handed neutrinos effectively decouple from the MSSM spectrum. Nevertheless, they can leave signals in LFV processes as we will see later. 

Reagarding the MSSM Higgs sector, both Higgs doublets, $H_u$, $H_d$ can be contained in the $4-2-2$ representation  $h=(1,2,\bar{2})$, allowing for Yukawa coupling unification  \cite{Gomez:2020gav}. However, in this work we consider a more general scenario where the MSSM Higgs fields are realized as superpositions of the appropriate Higgs fields residing in different representations mentioned in the $SO(10)$ and $4-2-2$ breakings. Hence, we relax the condition of Yukawa unification and keep free the ratio of the Higgs vev's, $tan\beta= v_u/v_d $. Also, we assume independent SSB mass terms for the MSSM Higgs fields.

Thus, the model below $\mgut$ turns out to be MSSM with some relations among the SSB parameters at the GUT scale: 
\begin{itemize}
    \item Gaugino masses. Although there is a $4-2-2$ realtion among gaugino masses, the identification $M_3=M_4$ and $M_2=M_{2L}$, together with breaking of the $L/R$ symmetry that implies $M_{2R}=y_{LR}\cdot M_{2L}$, results in independent MSSM gaugino masses. 
    \item Scalar masses. Common masses for squarks and leptons but distinction between the left and right fields $m_{R}\equiv x_{LR}\cdot m_{L}$. Different Higgs mass parameters: $m_{H_u}= x_u\cdot M_L$,  $m_{H_d}= x_d\cdot M_L$.   
\end{itemize}

To summarize, the $4-2-2$ GUT symmetry determines a MSSM with the following free parameters: 
\begin{equation}
    M_4, M_{2L}, y_{LR}, M_L, x_{LR}, x_{u}, x_{d}, A_0/M_L, tan\beta,
\label{eq:mssm-soft} 
\end{equation}
where the masses can go up to the TeV scale while the parameters $ y_{LR}$, $x_{LR}$, $x_{u}$, $x_{d}$ and $A_0/M_L$ are of order unity. Using the relations explained above, it is possible to identify the extracted MSSM framework with 

\begin{equation}
    M_3, M_2, M_1, M_L, M_R, m_{H_u}, m_{H_d}, A_0, tan\beta,
\label{eq:ps-soft}    
\end{equation}
as free parameters.

The model is complemented with a mechanism that can introduce LFV without interfering with the flavor conserving SUSY contributions to muon $g-2$ and DM studied in \cite{Gomez:2022qrb}. This can be done by assuming a see-saw mechanism that generates LFV slepton masses below the GUT scale, or additional U(1) family symmetries  that explain the flavor structure of the Yukawa couplings and are broken at a scale above the $M_{GUT}$ introducing family dependent soft sfermion  masses that still preserve the $4-2-2$ symmetry. In the resulting MSSM, the fermions and sfermions mass matrices are not aligned in the same superfield basis leading to a SUSY contribution to flavor violating processes in charged leptons \cite{Calibbi:2017uvl, Vicente:2015cka} and quarks, like B and kaon systems \cite{Isidori:2006qy,Altmannshofer:2009ne,Crivellin:2017gks, Endo:2017ums}, imposing strong bounds on the flavor structure of the models \cite{Olive:2008vv, Gomez:1998wj}. However, the RG induced gauino contribution to soft masses is flavor independent such that potential flavor mixing soft masses introduced at GUT can be substantially reduced at the EW scale 
\cite{Gomez:2010ga}. Furthermore, in the $4-2-2$ framework under consideration, despite the fact that squarks and sleptons get common SSB terms, the splitting of the gaugino masses induce squark masses much larger than the sleptons ones. Therefore, in this framework we can find significant SUSY contributions to LFV without  enhancing FV in the quark sector.

\subsection{SUSY contribution to muon \em{g-2}}
\label{subsec:g2}

The supersymmetric contributions at the leading one-loop level are shown in Fig. \ref{fig:SNloops}. The dominant contributions calculated by the mass insertion \cite{Moroi:1995yh,Stockinger:2006zn,Cho:2011rk, Fargnoli:2013zia} can be summarized as

\begin{equation}
\Delta a_{\mu}^{SUSY}\simeq (\Delta a_{\mu}^{\tilde{B}\tilde{\mu}_{L}\tilde{\mu}_{R}}+ \Delta a_{\mu}^{(\tilde{B}-\tilde{H})\tilde{\mu}_{L}}+ \Delta a_{\mu}^{(\tilde{H}-\tilde{B})\tilde{\mu}_{R}}+
\Delta a_{\mu}^{(\tilde{H}-\tilde{W})\tilde{\mu}_{L}}+
\Delta a_{\mu}^{(\tilde{W}-\tilde{H})\tilde{\nu}_{\mu}}),
\end{equation}
where the contribution from each diagram can be expressed as \cite{Cho:2011rk}: 
\begin{subequations}
\label{eq:g2_eq} 
\begin{eqnarray}
\Delta a_{\mu}^{\tilde{B}\tilde{\mu}_{L}\tilde{\mu}_{R}}\;\;\;
&=& 
 \phantom{-}  \frac{g_Y^{2}}{8 \pi^2}
\frac{m^2_\mu \mu \tan \beta}{M_1^3} \ 
F_b \left( \frac{m^2_{\tilde{\mu}_L}}{M_1^2}, 
           \frac{m^2_{\tilde{\mu}_R}}{M_1^2} \right),
\label{eq:N1} \\
 \Delta a_{\mu}^{(\tilde{H}-\tilde{B})\tilde{\mu}_{R}}
&=&
- \frac{g_Y^{2}}{8\pi^2}
 \frac{m^2_\mu M_1 \mu \tan \beta}{m^4_{\tilde{\mu}_R}} \
F_b \left( \frac{M_1^2}{m^2_{\tilde{\mu}_R}},
           \frac{\mu^2}{m^2_{\tilde{\mu}_R}} \right),
\label{eq:N2}  \\
\Delta a_{\mu}^{(\tilde{B}-\tilde{H})\tilde{\mu}_{L}}
&=&
\phantom{-} \frac{g_Y^{2}}{16\pi^2}
 \frac{m^2_\mu M_1 \mu \tan \beta}{m^4_{\tilde{\mu}_L}} \ 
F_b \left( \frac{M_1^2}{m^2_{\tilde{\mu}_L}}, 
           \frac{\mu^2}{m^2_{\tilde{\mu}_L}} \right), 
\label{eq:N3}  \\
\Delta a_{\mu}^{(\tilde{W}-\tilde{H})\tilde{\nu}_{\mu}}
&=& 
 - \frac{g^2}{16\pi^2}
 \frac{m^2_\mu M_2 \mu \tan \beta}{m^4_{\tilde{\mu}_L}} \
F_b \left( \frac{M_2^2}{m^2_{\tilde{\mu}_L}},
           \frac{\mu^2}{m^2_{\tilde{\mu}_L}} \right),
\label{eq:N4} \\
\Delta a_{\mu}^{(\tilde{W}-\tilde{H})\tilde{\nu}_{\mu}}
&=& 
 \phantom{-}  \frac{g^2}{8\pi^2}
\frac{m^2_\mu M_2 \mu \tan \beta}{m^4_{\tilde{\nu}}} \ 
F_a \left( \frac{M_2^2}{m^2_{\tilde{\nu}}}, 
           \frac{\mu^2}{m^2_{\tilde{\nu}}} \right). 
\label{eq:C2} 
\end{eqnarray}
\end{subequations}
whith, 
\begin{subequations}
\label{eq:loopfuction} 
\begin{eqnarray}
F_{a}(x,y) &=& -\dfrac{1}{2(x-y)}\left[\dfrac{(x-1)(x-3)+2 \ln(x)}{(x-1)^{3}} 
- \dfrac{(y-1)(y-3)+2 \ln(y)}{(y-1)^{3}}\right]~,\\
F_{b}(x,y) &=& -\dfrac{1}{2(x-y)}\left[\dfrac{(x-1)(x+1)+2 x\ln(x)}{(x-1)^{3}} 
- \dfrac{(y-1)(y+1)+2 y \ln(y)}{(y-1)^{3}}\right].
\end{eqnarray}
\end{subequations}

In the framework of the $4-2-2$ symmetry presented in Ref.~\cite{Gomez:2022qrb}, we find models that can provide simultaneous explanations for the measured muon $g-2$ and the desired DM candidate consistent with the Planck measurements \cite{Planck:2018nkj}.  Typically, the main SUSY contributions to muon $g-2$ in our models come from the processes represented with the top-left and bottom-right diagrams of Fig. \ref{fig:SNloops}, corresponding to Eqs.(\ref{eq:N1}) and (\ref{eq:C2}), respectively. The dominance of one or the other diagram can be associated with different features of the model,  such as the nature of LSP and its mass difference with the other SUSY particles that coannihilate with it in order to explain the observed DM abundance. All contributions are proportional to $\mu \tan\beta$, however, the value of this term is limited by RGE-induced electroweak symmetry breaking (EWSB) and the vacuum stability of the MSSM scalar potential~\cite{Carena:2012mw}.

\begin{table}[ht!]
\centering
\begin{tabular}{c|c}
SUSY Scale & GUT Scale \\ \hline
$m_{\tilde{\mu}_{L}}, m_{\tilde{\nu}}$ & $m_{L}$ \\ 
$m_{\tilde{\mu}_{R}}$ & $m_{R}$ \\
$M_{\tilde{B}}$ & $M_{1}$ \\
$M_{\tilde{W}}$ & $M_{2L}$ \\
$\mu$ & $m_{H_{u}}, m_{H_{d}}$ \\
$A_{\mu}$ & $A_{0}$ \\
$\tan\beta$ & $\tan\beta$ \\ \hline
\end{tabular}
\caption{\it{The fundamental parameters determining the supersymmetric contributions to muon $g-2$ at the low scale (left) and the GUT scale (right).}}
\label{tab:fund}
\end{table}

\subsection{LFV decays in SUSY \em{4-2-2}.}
\label{subsec:LFV}

 We will work within the  $4-2-2$ framework presented in  Ref.~\cite{Gomez:2022qrb}. The relevant SUSY parameters at the GUT scale are listed in Table~\ref{tab:fund}. As previously shown, in this framework the muon $g-2$ solutions favor heavy gluinos, while the LSP can still be a good DM candidate with masses lighter than about 500 GeV.  The heavy gluinos enhance the squark masses through RGEs, thus leading to a significant splitting between the squarks and slepton masses. This can easily be seen by using the approximate RGE solutions provided in Refs.~\cite{Ibanez:1984vq, Martin:1997ns} that show an approximate relation between the GUT scale parameters and the low energy mass terms. 
 Although the results presented in subsequent  sections are obtained by running the full RGE set by using the public code SPheno \cite{Porod:2003um, Porod:2011nf}, these approximations allow us to understand the relations amoung SUSY masses in 
 the different models and thus, the relevance of the SUSY contributions to the processes under consideration.
%The values of the gaugino masses are:
%\begin{equation}
%M_{\tilde{B}} \sim 0.41 M_1,\qquad M_{\tilde{W}} \sim 
%0.67 M_2 ~, 
%\label{eq:bino}
%\end{equation}
 
\begin{eqnarray}
m_{\tilde{Q}_i}^2\!\!\!&=&\!\!\! m_L^2 + K_3 + K_2 + {\frac{1}{36}}K_1,
\label{mqL} \nonumber \\
m_{\tilde{U}_i}^2 \!\!\! &=&\!\!\! m_R^2 + K_3
\qquad\>\>\>
+ {\frac{4}{9}} K_1,
 \nonumber \\
m_{\tilde{D}_i}^2  \!\!\!&=&\!\!\! m_R^2 + K_3
\qquad\>\>\>
+ {\frac{1}{9}} K_1,
 \nonumber \\
m_{\tilde{e}_{Li}}^2 \!\!\!&=&\!\!\! m_L^2 \qquad\>\>\> + K_2 + {\frac{1}{4}} K_1,
 \nonumber \\
m_{\tilde{e}_{Ri}}^2 \!\!\!&=&\!\!\! m_R^2
\qquad\qquad\>\>\>\>\>\> \, +\, K_1.
\label{eq:sfermions}
\end{eqnarray} 
The quantities $K_3$, $K_2$ and $K_1$ from the RG running proportional to the gaugino masses. Explicitly, they are found at one loop order by solving the RGEs, as given for example in Ref.~\cite{Martin:1997ns} :
\begin{equation}
K_i(Q) = \begin{Bmatrix}
{3/5}&\\
{3/4}&\\
{4/3}\end{Bmatrix} 	
 \times
\frac{1}{2 \pi^2} \int^{{\rm ln} M_{GUT}}_{{\rm ln}Q }dt\>\,
g^2_a(t) \,|M_i(t)|^2\qquad\>\>\> (i=1,2,3),
\label{eq:ks}
\end{equation}
where the running parameters $g_i(Q)$ and $M_i(Q)$ obey the RGEs given in Ref.~\cite{Martin:1997ns}.With the values of $Q=M_{SUSY}=\sqrt{m_{\tilde t_1}m_{\tilde t_2}}$ that we find in our models, the ranges of these amounts are:
\begin{equation}
K_1 \approx 0.15 \times (M_1^{GUT}) ^2,\quad
K_2 \approx (0.3-0.5) \times (M_2^{GUT}) ^2,\quad
K_3 \approx (4.3- 6.7) \times (M_3^{GUT}) ^2.
\label{k123}
\end{equation}
Although this approximation is only valid for cases where first and second generation Yukawa couplings can be neglected still it provides a good picture of the mass splitting among sqarks and sfermions. Furthermore, these equations can show how a possible flavor dependence of the scalar soft terms can be transmitted to the low energy observables. For instance, we observe that the sfermion masses at the TeV scale have a  flavor independent component arising from gauginos in addition to the contribution from the soft scalar masses, that can be flavor dependent. This may induce misalignments between the fermion Yukawa and the sfermion mass matrices. 

Typically, in SUSY models the flavor mixture of the soft terms results in fermionic flavor violating decays, heavily constrained by the strong experimental bounds~\cite{Gomez:1998wj,King:2004tx,Olive:2008vv}. However, in the $4-2-2$ framework we find that the large gluino masses suitable for explaining muon $g-2$, and the lower masses for $M_1, M_2$ that can explain DM, helps suppress such processes in the quark sector while being less restrictive in the leptonic sector. For instance, although all the sfermions may get FV masses of the same order at the GUT scale, according to  Eqs.~\ref{eq:sfermions} only the diagonal mass terms grow with gaugino masses, and thus the flavour mixing in the squark mass matrices are washed out by the large gluino contribution to the diagonal ones. Therefore, the SUSY contribution to sensitive systems like kaon decays \cite{Aebischer:2022vky}  is not significant. On the other hand, the moderate values we find for the lighter gauginos do not suppress the effect of the slepton mass mixing terms, leading to the  prediction of charged lepton violating (cLFV)  decays on the range of the experimental searches~\cite{Gomez:2010ga}.

Flavor dependent scalar soft terms can be induced below the GUT scale through the RGEs when a mechanism like the see-saw is present in the theory to explain neutrino masses. Also, the flavor dependence can be generated above the GUT scale if the GUT symmetry is combined with additional ones to explain the favour hierarchy in the Yukawa interactions. 

In the current work, we use gauge group $4-2-2$ without imposing Left-Right (LR) symmetry (for a detailed description see \cite{Gomez:2020gav,Gomez:2022qrb} and references therein). Such symmetries can emerge from $SO(10)$ breaking by the Higgs fields in $210_{H}$ representation \cite{Babu:2016bmy}. Assuming  I see-saw, cLFV will be induced below the GUT scale. The running of the slepton masses from $M_{GUT}$ to $M_R$ is affected by the Dirac neutrino Yukawa coupling matrix $Y_{\nu}$, that can be of the same order as the fermion Yukawa couplings. Therefore, flavor-changing terms for the slepton masses are introduced since $Y_\nu$ cannot be diagonalised in the same basis as the charged lepton Yukawa matrix $Y_l$. For instance, in a basis where $Y_l$ is diagonal at the leading-log approximation~\cite{Hisano:1995cp}, the non-diagonal terms of the lepton mass matrix take the following form at scale $M_R$: 

\begin{eqnarray}
(m_{\tilde L}^2)_{ij} &\sim & \frac 1{16\pi^2} (6m^2_0 + 2A^2_0)
\left({Y_{\nu}}^{\dagger} Y_{\nu}\right)_{ij}  
\log \left( \frac{M_{\rm GUT}}{M_R} \right) \, , \nonumber\\
(m_{\tilde e}^2)_{ij} &\sim & 0  \, , \nonumber\\
(A_l)_{ij} &\sim & \frac 3{8\pi^2} {A_0 Y_{l}}_i
\left({Y_{\nu}}^{\dagger} Y_{\nu}\right)_{ij}  
\log \left( \frac{M_{\rm GUT}}{M_R} \right) \, .
\label{eq:see-saw}
\end{eqnarray}

  In addition, to explain fermion flavour it is possible to combine family symmetries with the GUT group to build predictive Yukawa textures \cite{Ibanez:1994ig,Lola:1999un,Ellis:1998rj,Kane:2005va,King:2000ge,Dent:2004dn}. In some of these frameworks,  the appropriate hierarchy among couplings can be reproduced with a single parameter obtained through the Froggatt-Nielsen mechanism \cite{Froggatt:1978nt}. Moreover, in models with soft SUSY breaking terms derived from supergravity, flavor dependent scalar soft terms can be induced above the GUT scale when the unifying symmetry is combined with  additional groups under which the fermion families carry different charges.  Consider, for instance, higher order operators at some high scale $M$:

\begin{equation}
    \Phi_i \Phi_J^c h\; \frac{\theta^{(q_i-q_j+ q_h)}}{M}, 
\end{equation}
where $\Phi_i$ and $\Phi_j^c$ ($i,j=1,2,3$) denote quark/lepton superfields, h is a Higgs superfield, and $\theta$ represents the flavon field, which develops a non-zero vacuum expectation value (VEV) such that a small expansion parameter is induced, $\epsilon = \dfrac{\langle \theta \rangle}{M}$, where $\langle \theta \rangle$ is the flavon VEV. If the flavor symmetry is chosen to be an abelian  $U(1)$ group, the hierarchy among the flavors can arise from non-universal charges of the families under this group. Even if flavor-independent soft terms are assumed above the GUT scale, the soft terms generated at $M_{{\rm GUT}}$ are flavor-dependent. Their flavor structure depends on the family charges, normalization of K{\"ahler} potential, and the rotation of the Yukawa matrices \cite{King:2004tx, Das:2016czs}. Therefore, in a field basis where the fermion mass matrices are diagonal, the corresponding sfermion matrix takes the form: 

\begin{eqnarray}
m_f^2=
 \begin{pmatrix}
1 &  \epsilon^{q^{12}}&  \epsilon^{q^{13}}\\
\epsilon^{q^{12}}&  1& \epsilon^{q^{23}}\\
\epsilon^{q^{12}} & \epsilon^{q^{23}} & 1\\
\end{pmatrix}
\times m_{f^0}^2, 
\label{eq:msoft}
\end{eqnarray}
where $q^{i,j}$ are combinations of charges of the generations $i, j=1,2,3$, and $m_{f^0}$ is a common soft SUSY breaking mass. 

Due to the lack of LR symmetry, the sfermions mass matrices are independently induced for the left- and right-handed fields, and their flavor-dependent forms can be written as follows:

\begin{equation}
m_f^2=
 \begin{pmatrix}
1 &  \epsilon^{q_L^{12}}&  \epsilon^{q_L^{13}}\\
\epsilon^{q_L^{12}}&  1& \epsilon^{q_L^{23}}\\
\epsilon^{q_L^{12}} & \epsilon^{q_L^{23}} & 1\\
\end{pmatrix}
\times m_L^2, \hspace{0.5cm}
m_{f^c}^2=
 \begin{pmatrix}
1 &  \epsilon^{q_R^{12}}&  \epsilon^{q_R^{13}}\\
\epsilon^{q_R^{12}}&  1& \epsilon^{q_R^{23}}\\
\epsilon^{q_R^{12}} & \epsilon^{q_R^{23}} & 1\\
\end{pmatrix}
\times m_R^2,
\label{eq:mFmFc}
\end{equation}

\noindent
where $m_{L}$ and $m_{R}$ are the soft sypersymmetry breaking (SSB) masses for the left- and right-handed sfermions. These matrices are induced both in the squark the slepton sectors. However, because of the heavy squarks driven by the heavy gluinos through RGEs, the flavor mixing in these matrices is suppressed and their flavor mixing in essentially coincides with the CKM matrix. Therefore, sensitive processes like rare kaon decays \cite{Aebischer:2022vky} are not significantly affected in this models. On the other hand, even though the charged lepton Yukawa matrix remains diagonal in this basis, the non-diagonal form of the slepton mass-squared matrix can induce the cLFV processes such as $l_{i}\rightarrow l_{j}\gamma$ at the loop level, as shown diagrammatically in Fig. \ref{fig:muegamma}.

 In summary we classify the LFV  scenarios according to the scale where mixing of the soft terms is introduced: 

\begin{itemize}
    \item {\bf LFV below the GUT scale.}  In this case the soft terms are taken to be  universal at $M_{GUT}$, and LFV is induced through a type I  see-saw mechanism We use the simplified scenario presented in Ref.~\cite{Ellis:2020jfc}, where a simple but representative case was obtained by considering a common RH neutrino mass $M_R$ is assumed for the three species. In this case, using \cite{Casas:2001sr}, we find for the see-saw parameters that enter eq.~(\ref{eq:see-saw}):
\begin{equation}
Y_\nu^\dagger Y_\nu= \frac{2}{v_u^2}M_R U m_\nu^\delta U^\dagger~.
\label{eq:ynu2}
\end{equation}
Using a basis in which the charged lepton Yukawa matrix is diagonal, 
$U$ can be identified with the Pontecorvo-Maki-Nakagawa-Sakata (PMNS) matrix, while $m_\nu^\delta$ is the diagonal neutrino mass matrix. $U$ is set to the central values of the experimental data \cite{ParticleDataGroup:2022pth}, while the values $m_\nu^\delta=Diag(1.1\cdot 10^{-3}, 8 \cdot 10^{-3}, 5 \cdot 10^{-2})$~eV are compatible with the observed neutrino oscillations, assuming direct mass hierarchy. With the assumption of common masses for the heavy Majorana neutrinos, the LFV effects are independent of textures for the matrices $M_R$ and $Y_\nu$, which in some cases can lead to some cancellations in the LFV branching ratios~\cite{Cannoni:2013gq}. However, these ratios are strongly dependent on $M_R$. Here, we assume $M_R= 2.5 \times 10^{12}$ GeV and,  with this choice, we obtain values for BR($\mu \rightarrow e \gamma$) of the order of the current or projected experimental bounds. 

\item{\bf Family symmetries with flavor dependent soft terms at  $M_{GUT}$}. In this case,  we assume flavor mixing among the charged leptons from Eqs.~(\ref{eq:mFmFc}). We consider two separate  cases, namely  weak flavor mixing ($\epsilon=0.05$), and strong flavor mixing ($\epsilon = 0.2$). In our computations we select $q_L^{12}=q_L^{13}=3$, $q_L^{23}=2$ and $q_R^{12}=4$, $q_R^{13}=q_L^{23}=2$. These values, together with $\epsilon$, provide predictions for BR($\mu\rightarrow e \gamma$)  that are compatible with both the experimental data and the results we obtain from the see-saw model. We should remark that our choice of charges and values for $\epsilon$ is not aimed at a  full  explanation of the hierarchy of the Yukawa coupling matrices that may require, in addition to the family symmetries, a complex Higgs potential that splits quarks and leptons from the L and R multiplets in $4-2-2$ \cite{King:2000ge,Dent:2004dn}. Also, we neglect in this case the possible see-saw effects which can be consistent with the assumption of right-handed neutrinos masses $M_{N} \lesssim \mathcal{O}(10^{10})$~ GeV, accompanied by very small Dirac Yukawa couplings between the Higgs boson and neutrinos ($Y_{\nu} \lesssim 10^{-7}$) \cite{Coriano:2014wxa,Khalil:2010zza,Abbas:2007ag,Un:2016hji}, leading to tiny contributions to LFV processes.

\end{itemize}

\section{Fundamental Parameters and Experimental  Constraints}
\label{sec:scan}
\begin{comment}
    
We perform random scans in the fundmental parameter space of 4-2-2 with broken LR symmetry by using SPheno-4.0.4 \cite{Porod:2003um,Porod:2011nf}, which is implemented by SARAH-4.14.4 \cite{Staub:2008uz}. 
\end{comment}

In \cite{Gomez:2022qrb} we have shown the existence of models with soft terms derived from the $4-2-2$ model that can simultaneously explain  muon $g-2$ without running into conflict with the observed DM relic density. The input parameters at the GUT level for these models correspond to values of the $4-2-2$ parameters in eq.~\ref{eq:ps-soft} as follows:
\begin{equation}
\begin{array}{lll}
0 \leq & m_{L} & \leq 5 ~{\rm TeV} \\
0 \leq & M_{2L} & \leq 5 ~{\rm TeV} \\
-3 \leq & M_{3} & \leq 5 ~{\rm TeV} \\
-3 \leq & A_{0}/m_{L} & \leq 3 \\
1.2 \leq & \tan\beta & \leq 60 \\
0\leq & {\rm x}_{{\rm LR}} & \leq 3 \\
-3 \leq & {\rm y}_{{\rm LR}} & \leq 3 \\
0 \leq & {\rm x}_{{\rm d}} & \leq 3 \\
-1 \leq & {\rm x}_{{\rm u}} & \leq 2~. \\
\end{array}
\label{eq:paramSpacePSLR}
\end{equation}

The low energy spectrum and phenomenological constraints are obtained using SPheno-4.0.4 \cite{Porod:2003um,Porod:2011nf}, which is implemented by SARAH-4.14.4 \cite{Staub:2008uz}. This is done by using the MSSM parameters listed in eq.~\ref{eq:mssm-soft} and using MSSM and the see-saw type I packages included in the SPheno distribution. The selected models are compatible with EWSB and satisfy the GUT unification condition as described in  \cite{Gomez:2022qrb}. In addition, we vary the universal trilinear scalar interaction term $A_{0}$ by {requiring that} the magnitude of its ratio to $m_{L}$ {is not} greater than 3, to avoid the color/charge breaking minima of the scalar potential \cite{Ellwanger:1999bv,Camargo-Molina:2013qva,Camargo-Molina:2013sta}. In scanning the parameter space we use the Metropolis-Hastings algorithm \cite{Belanger:2009ti,Baer:2008jn}, and generate the solutions by following flat priors \cite{Trotta:2008bp}.

In our search, we keep the points that satisfy the constraints: 
\begin{equation}
	%\setstretch{1.8}
	\begin{array}{c}
		m_h  = 123-127~{\rm GeV}~,\\
		m_{\tilde{g}} \geq 2.1~{\rm TeV}~(800~{\rm GeV}~{\rm if~it~is~NLSP})~,\\
		0.8\times 10^{-9} \leq{\rm BR}(B_s \rightarrow \mu^+ \mu^-) \leq 6.2 \times10^{-9} \;(2\sigma)~, \\
		2.99 \times 10^{-4} \leq  {\rm BR}(B \rightarrow X_{s} \gamma)  \leq 3.87 \times 10^{-4} \; (2\sigma)~.
		% \\
%		0.114 \leq \Omega_{{\rm CDM}}h^{2} \leq 0.126~.
		\label{eq:constraints}
	\end{array}
\end{equation}
The Standard Model (SM) Higgs boson mass is computed by SPheno which runs 
three-loop RGEs by considering the effective SM Higgs potential
between $M_Z$ and $M_{SUSY}$ and imposing the two-scale matching condition at $M_{SUSY}$~\cite{Staub:2017jnp}. The SPheno outputs are transferred
to micrOMEGAs \cite{Belanger:2020gnr}, which calculates the relic density and
the scattering cross-sections of the DM candidates. We consider the 5-$\sigma$ Planck range~\cite{Planck:2015fie}:
\begin{equation}
 0.114 \leq \Omega_{{\rm CDM}}h^{2} \leq 0.126~.
 \label{eq:Planck}
\end{equation}
In general, we keep models that are not in contradiction with these bounds. That is, models that predict DM relic density below the upper bound in the above equation. However, we will present some of our results for the subsets of models predicting DM predictions within the Planck range. 
%		0.114 \leq \Omega_{{\rm CDM}}h^{2} \leq 0.126~.

In contrast to the study presented in \cite{Gomez:2022qrb}, here we perform a more selective  scan accepting only models that explain muon $g-2$ at $2\sigma$. The discrepancy between the combined BNL
and Fermilab results \cite{Muong-2:2024hpx}  and the data-driven theoretical value is \cite{Aoyama:2020ynm}: 
\begin{equation}
\Delta a_\mu = (24.9 \pm 4.8) \times 10^{−10} .
\end{equation} The impact on the SUSY models from   more restrictive ranges, for instance like the ones arising from lattice calculations, are studied in detail in Ref.\cite{Ellis:2024ijt}.

As indicated in the previous section, we introduce LFV in three different scenarios. One is a simplified seesaw type I, where the flavor mixing among left sleptons is induced by the RGEs. The other two scenarios mix the slepton flavors at scales above $\mgut$. The texture is chosen so that, for $\epsilon=0.05$, it provides values for BR($\mu\rightarrow e\gamma$) comparable to the seesaw case, while for $\epsilon=0.2$ it may strongly restrict the GUT values of the soft masses, allowing only models comparable with "no-scale" SUSY \cite{Lahanas:1986uc,Ellis:2013nka,Li:2016bww,Ford:2019kzv,Li:2021cte,Ellis:2020mno}. We impose the  the bounds \cite{ParticleDataGroup:2022pth}:
\begin{eqnarray}
	{\rm BR}(\mu\rightarrow e \gamma)&<& 4.2 \times 10^{-13}~,\\
	{\rm BR}(\tau\rightarrow e \gamma)&<& 4.3 \times 10^{-8}~,\\
	{\rm BR}(\tau\rightarrow \mu \gamma)&<& 3.2 \times 10^{-8}.
\end{eqnarray}

The DM constraints together with the muon $g-2$ condition and the cLFV bounds impose a severe restriction on the possible DM scenarios, that are limited to the following cases: 
\begin{itemize}

\begin{figure}[ht!]
\centering
\begin{tabular}{ccc}
\includegraphics[scale=0.26]{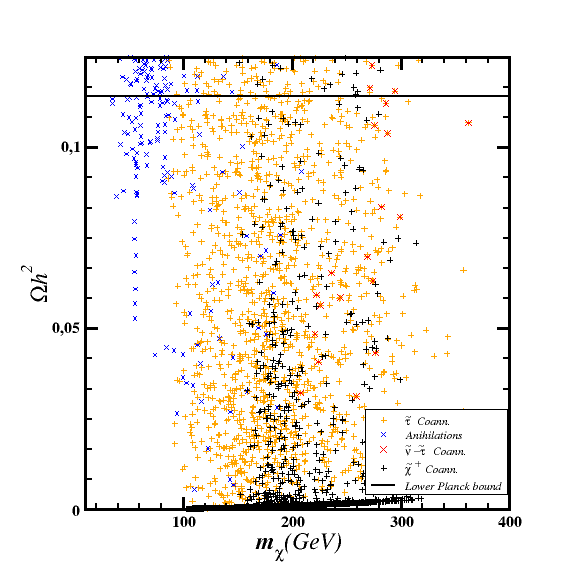}
\hspace{-.4 true cm}
\includegraphics[scale=0.26]{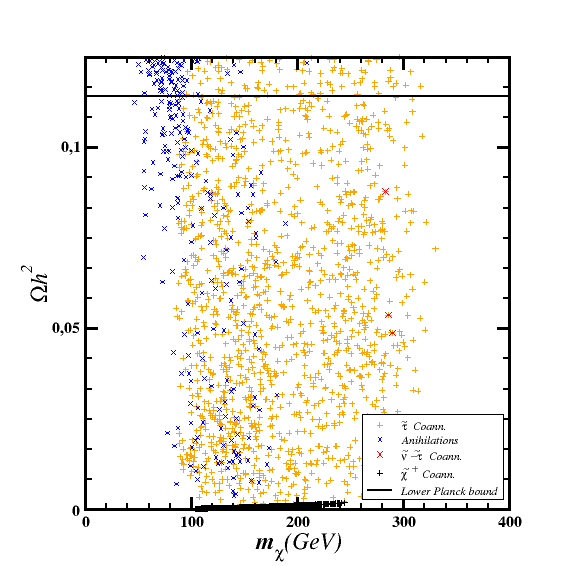}
\hspace{-.4true cm}
\includegraphics[scale=0.26]{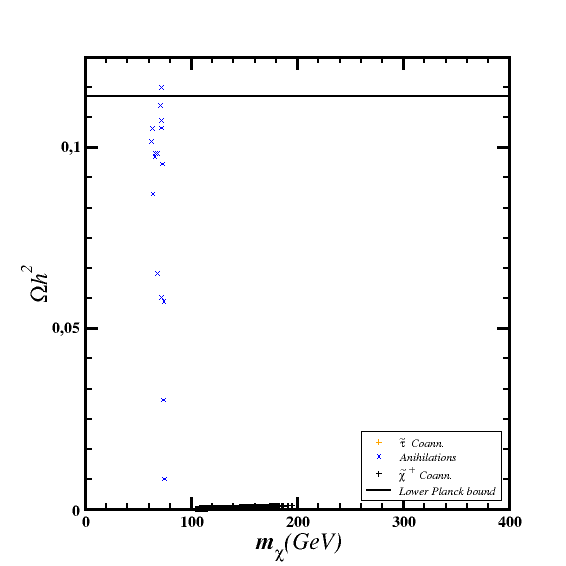}
%\hspace{-0.5cm}
\end{tabular}
\caption{\it Values of $\Omega h^2$ vs $m_\chi$ for models satisfying the Planck upper bound on DM and the  muon $g-2$ within $2\sigma$ in correlation with $m_\chi$. All the models predict charged lepton violating decays inside the experimental bounds. Different symbols and color codes are assigned to each class of model: blue crosses for annihilations,  orange pluses for $\tilde{\tau}-\chi$ coannihilations, red crosses for $\tilde{\nu}-\tilde{\tau}-\chi$ coannihilations, black pluses for $\chi^\pm-\chi$ coannihilations. The left panel corresponds to the see-saw case, while the two others correspond to scenarios with U(1) family symmetries. The centre one uses $\epsilon=0.05$ as expansion parameter while the right one uses $\epsilon=0.2$.}
\label{fig:oh2}
\end{figure}

\item[$\bullet$]{\bf $\chi$-annihilations.}
This scenario implies low mass values for the SUSY partners involved in the  muon $g-2$ computations. The LSP is mostly a Bino with annihilation cross section large enough to explain DM. The lightest sleptons are mostly right-handed. 

\item[$\bullet$]{\bf $\chi-\tilde{\tau}$-coannihilations.} 
This scenario implies a wider spectrum than the previous one. The LSP is also a Bino but, in this case, models explaining the muon $g-2$ require low values for mostly right sleptons, that also decrease the DM abundance to Planck levels.

\item[$\bullet$]{\bf $\chi-\tilde{\tau}-\tilde{\nu}$-coannihilations.} 
This scenario is enabled by $L/R$-symmetry breaking, which allows for the possibility of left sleptons lighter than the right ones. In this scenario, the light $\tilde{\nu}$ becomes the NLSP that enters the coannihilation scenario and also increases the contribution to muon $g-2$. 

\item[$\bullet$]{\bf $\chi-\tilde{\chi}^\pm$-coannihilations.}
This scenario is allowed because of the gaugino mass relations introduced by the 4-2-2 symmetry. In this case, the LSP is a Wino while the lightest chargino is the NLSP. Values of the chargino masses are close to neutralinos and sleptons allow for relatively large values for muon $g-2$. However, many of the models predict relic densities below the Planck bounds. Therefore, the models, while cosmological viable, cannot resolve the DM problem. 
\end{itemize}

Fig.~\ref{fig:oh2} displays the values of $\Omega_\chi h^2$ vs the LSP mass for the three different scenarios under consideration. Each of the coannihilation scenarios listed above is shown with different colors and symbols as given in legends. We observe that a particle coannihilates with the LSP when its mass is up to 30\% larger than the LSP. In many cases, there are other particles in this range of mass in addition to the NLSP. We find models with $m_\chi$ below 100 GeV that satisfy the Planck bounds without coannihilations.  Such solutions represent the annihilation of LSP into the SM particles.

Different DM areas can be associated with the dominant contributions to the muon $g-2$. In models with Bino-like LSP, the dominant contribution (eq.~\ref{eq:N1}) arises from the upper left diagram of Fig.~\ref{fig:SNloops}, while the lower right diagram is the dominant one in scenarios with Wino-like LSP (eq.~\ref{eq:C2}). The contribution from this diagram is also enhanced by a light $\tilde{\nu}_\mu$. The LR asymmetry of the soft terms, together with $M_2>M_1$, allows the left sleptons to become lighter than the right ones. This selection of soft terms favors both SUSY contributions to muon $g-2$ and a relic density of LSP below the upper Planck bound. However, the small mass differences between the lightest SUSY particles produce significant changes in $\Omega h^2$. Indeed, models in Fig.~\ref{fig:M2M1} change from Bino to Wino LSP as $M_1$ increases. In regions with $M_1\sim M_2$, we obtain a zone between models with $\tilde{\tau}$ (orange points) or chargino NLSP (black crosses), where $\tilde{\nu}_\tau$ becomes the NLSP (red points). The nature of the LSP also plays an important role in DM implications. In the solutions with Wino-like LSP, we obtain a zone between models with $\tilde{\tau}$ (orange points) or chargino NLSP (black crosses), where $\tilde{\nu}_\tau$ becomes the NLSP (red points). Such solutions can be associated with $M_{1}\sim M_{2}$.

%Many of these models are s-till consistent with the LHC bounds because even if the mass difference with the NLSP is too large to coannihilate with the LSP, it is small enough to escape the LHC reach. The displayed models predict cLFV bounds inside current bounds, so that the number of models decreases as we use scenarios with larger mixing of the soft terms, as we see in the next subsection.

\section{Analysis and Results}
\label{sec:results}
\subsection{Slepton masses and LFV}

As previously mentioned, flavor violation is induced through the flavor-dependent mass matrices, but the large values of $M_{3}$ (see Fig.~\ref{fig:M3M1}) drives the diagonal entries to be quite large in the squark mass-squared matrices. Thus, the flavor violating entries do not have a strong impact in the quark sector. On the other hand, the desired muon $g-2$ values require moderate masses for the sleptons and electroweak gauginos, which implies $M_{1},M_{2} \ll M_{3}$. In this context, the GUT scale mass-squared matrices with the flavor violating entries can have considerable effects in the slepton mass spectrum, especially in the solutions leading to slepton-neutralino coannihilation processes.

\begin{figure}[!ht]
\centering
\begin{tabular}{ccc}
        \includegraphics[scale=0.26]{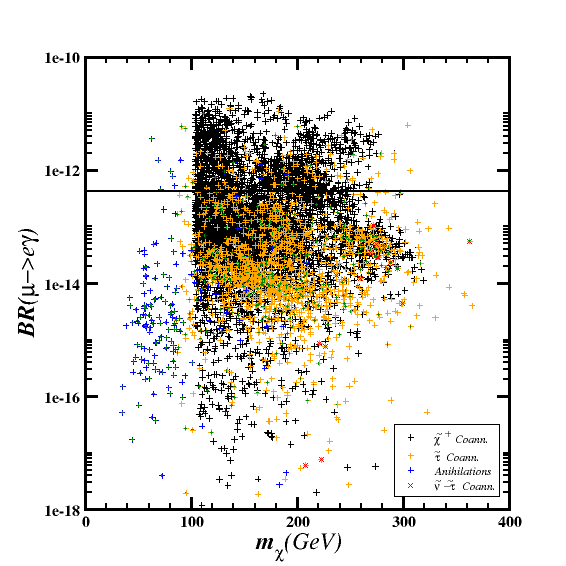}\hspace{-.4 true cm}
        \includegraphics[scale=0.26]{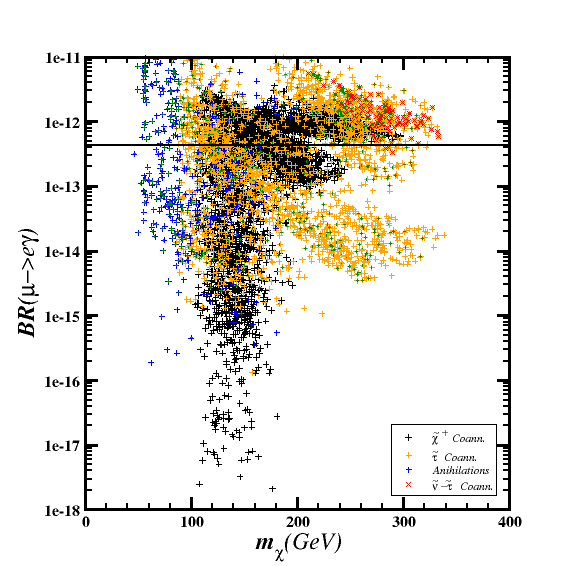}\hspace{-.4 true cm}
        \includegraphics[scale=0.26]{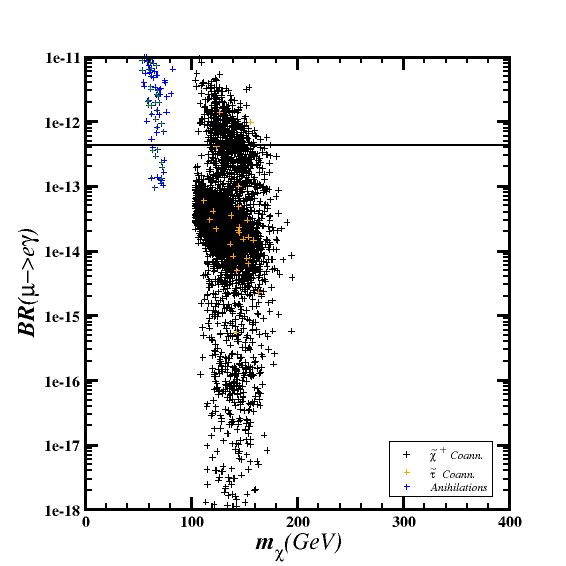}
\end{tabular}        a
\caption{\footnotesize BR($\mu\rightarrow e\gamma$) vs $m_\chi$ for the three models discussed in the text. Order of the panels, symbols and color codes are as in Fig.~\ref{fig:oh2}. In all the models the prediction for  $\Omega h^2$ is below the upper limit of eq.~\cite{Planck:2015fie}, the subset of points that are also above the lower bound are marked in green.}
\label{fig:mueg}
\end{figure}

\begin{figure}[!ht]
\centering
\begin{tabular}{ccc}
\includegraphics[scale=0.26]{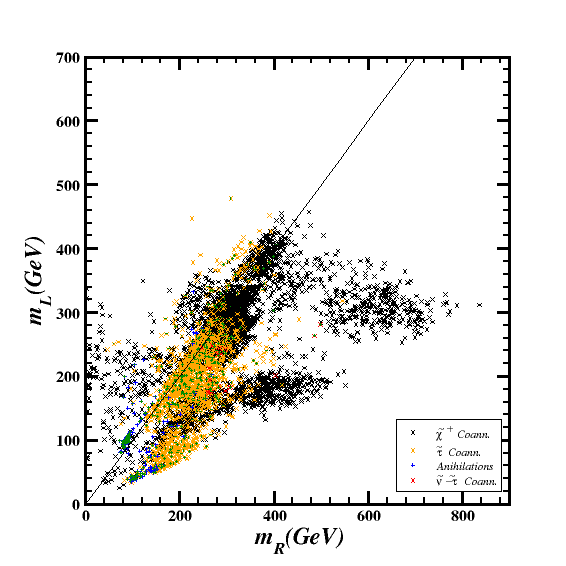}\hspace{-.4true cm}
        \includegraphics[scale=0.26]{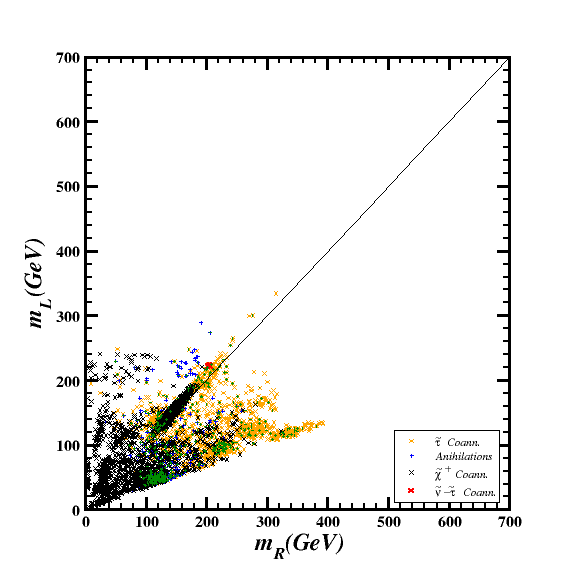}     \hspace{-.4true cm}
        \includegraphics[scale=0.26]{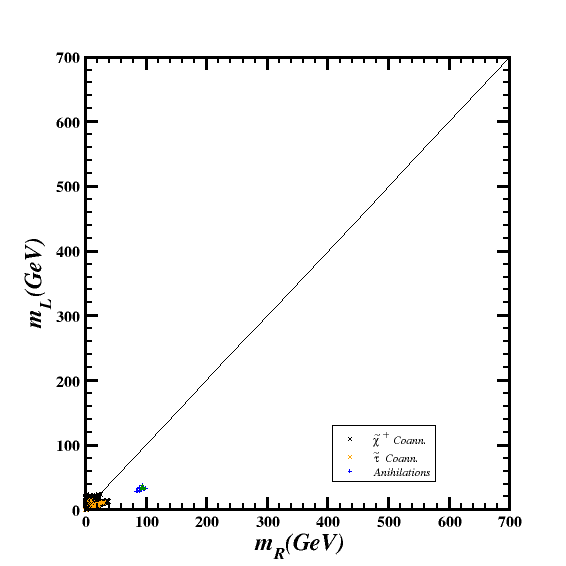}
\end{tabular}  
\caption{\footnotesize GUT scale values of $m_L$ vesus $m_R$. Order of panels, symbols and colors  are as in Fig.~\ref{fig:oh2}. All the points satisfy the BR($\mu\rightarrow e\gamma$) constraint. In all the models the prediction for  $\Omega h^2$ is below the upper limit of eq.~\cite{Planck:2015fie}, the subset of points that are also above the lower bound are marked in green.}
\label{fig:MLMR}
\end{figure}   

The predictions for BR($\mu\rightarrow e\gamma$) vs $m_\chi$ are displayed in Fig.~\ref{fig:mueg} for the three mixing scenarios under consideration. Confronting the results for $\mu\rightarrow e\gamma$ with the scales of $m_{L}$ and $m_{R}$ shown in Fig.~\ref{fig:MLMR}, we can summarize our findings as follows:

\begin{itemize}

\item In the seesaw scenario the constraint from $\mu \rightarrow e\gamma$ is very restrictive for the solutions with $m_{L} \gtrsim 100$ GeV, while it is less so for larger $m_{R}$ values. In contrast to the other scenarios, where the LFV terms are induced at $\mgut$, solutions with low values of $m_{L/R}$ are  not allowed in the seesaw scenario. The LEP limit on masses of the charged particles leads to a lower bound on $m_{L}$ and $m_{R}$ of about 20 GeV due to the light staus.

\item In the scenario with $\epsilon = 0.05$ (middle planes in Figs. \ref{fig:mueg} and \ref{fig:MLMR}), the $\mu \rightarrow e\gamma$ constraint restricts the scales for the sfermion masses as $m_{L} \lesssim 90$ GeV and $m_{R}\lesssim 50$ GeV, while beyond these limits the LFV processes become significant. In the region where the relic density is inside the Planck bound, the prediction for ${\rm BR}(\mu\rightarrow e\gamma)$ is one or two orders of magnitude lower than the experimental limit. The LSP in these regions is mostly Bino-like, and the desired DM relic density can be realized through stau-nuetralino coannihilation and annihilations of LSP pairs, while models with chargino-neutralino coannihilations predict a relic DM density below the Planck bound. 

\item The third scenario with $\epsilon = 0.2$ is obviously affected more by the LFV constraints (left panel of Fig.~\ref{fig:muegamma}). Comparing with the left panel of Fig.~\ref{fig:MLMR}, the only solutions consistent with the LFV constraints lie in the regions where $m_{L}\lesssim 15$ GeV and $m_{R} \lesssim 40$ GeV. The LSP neutralino relic density is reduced to the required ranges through the chargino-neutralino coannihilations in these solutions. Even though the LFV constraints are weaker for large values of $m_{R}$, it is also possible to realize some solutions consistent with the LFV constraints and even DM observations in the regions where $m_{L} < m_{R} \sim 100$ GeV.

\end{itemize}

In sum, the $\mu\rightarrow e\gamma$ constraint seems more restrictive in models where flavor mixing is introduced above $\mgut$. The seesaw mechanism induces LFV only in the left-handed sleptons while the scenarios with non-zero $\epsilon$ generate flavor mixing in both the left- and right-handed sfermions, thus enhancing the rates for the flavor violating processes. This can be potentially dangerous for decays such as $b\rightarrow s\gamma$ \cite{Olive:2008vv}. Even though this enhancement can be suppressed by heavy masses as in the case of squarks, the condition of muon $g-2$ solution within $2\sigma$ does not allow heavy masses for sleptons and electroweak gauginos. In the mass scales favored by the muon $g-2$ solution, the cLFV rates are enhanced by the masses of sleptons, and as a result, we observe a stronger impact on $m_{L}$ and $m_{R}$ from cLFV constraints in these two scenarios. 

\begin{figure}[!ht]
   \centering
%   \hspace{-0.7cm}
\begin{tabular}{ccc}
        \includegraphics[scale=0.26]{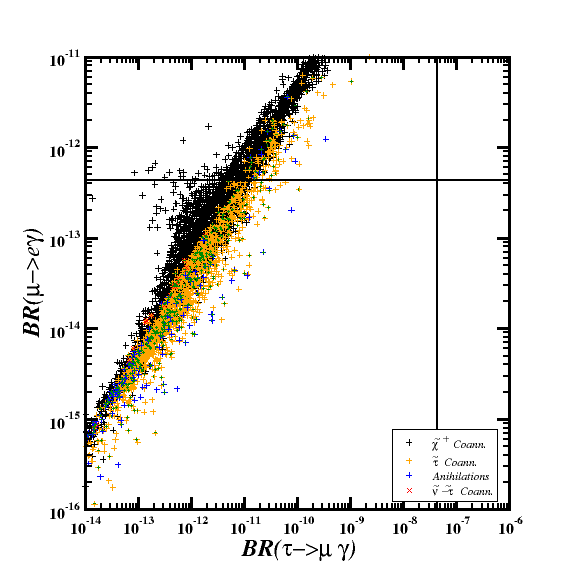}\hspace{-.4 true cm}
        \includegraphics[scale=0.26]{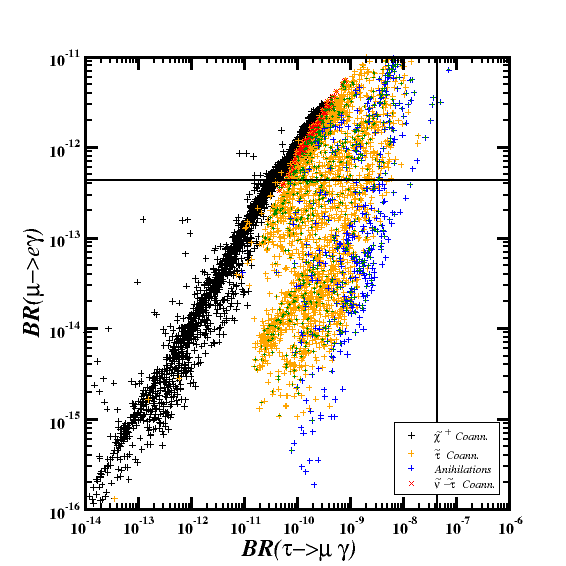}\hspace{-.4 true cm}
        \includegraphics[scale=0.26]{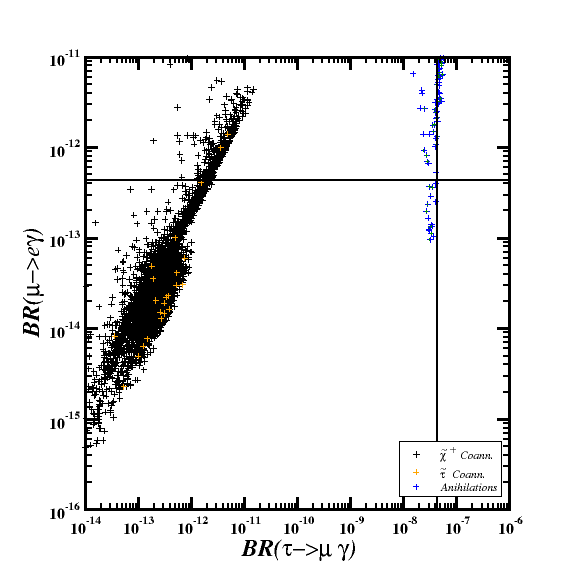}
        \end{tabular}
\caption{\footnotesize BR($\mu\rightarrow e\gamma$) vs.  BR($\tau\rightarrow \mu\gamma$), the panles,  color codes and symbols represent the same models as in the previous Fig.\ref{fig:mueg}. }
\label{fig:tme}
\end{figure}  

Another important observation is the correlation of the BR($\mu\rightarrow e\gamma$) with other cLFV decays, such as $\tau\rightarrow \mu\gamma$. Generic see-saw models predict more or less the same rate for these decays, while the experimental bounds differ by several orders of magnitude. On the other hand, in general, the prediction for flavor mixing related to the family symmetry can allow a different hierarchy for the LFV $\tau$ and $\mu$ decays. In Fig.~\ref{fig:tme} we display BR($\mu\rightarrow e\gamma$) vs BR($\tau\rightarrow \mu\gamma$), and observe that scenarios with family symmetries can predict branching ratios for these decays in ranges closer to the experimental values. In addition to fitting the models with the current cLFV constraints, the solutions will be potentially tested soon by the improvements and upgrades in MEG II~\cite{MEGII:2018kmf,MEGII:2023ltw,MEGII:2023fog}, Mu2e \cite{Hedges:2022tnh}, COMET \cite{COMET:2018auw} and DeeMe \cite{Teshima:2019orf}.

\subsection{Muon $g-2$ and SUSY Masses}
\label{subsec:g2SUSYmasses}
\begin{figure}[!ht]
\begin{tabular}{ccc}
\includegraphics[scale=0.26]{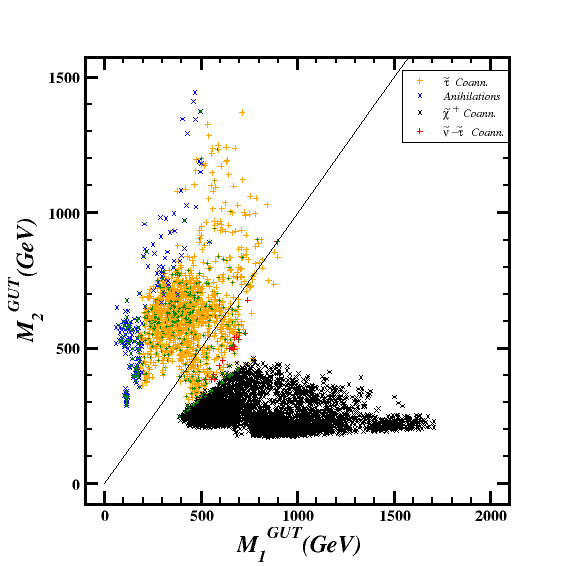}\hspace{-.4true cm}
        \includegraphics[scale=0.26]{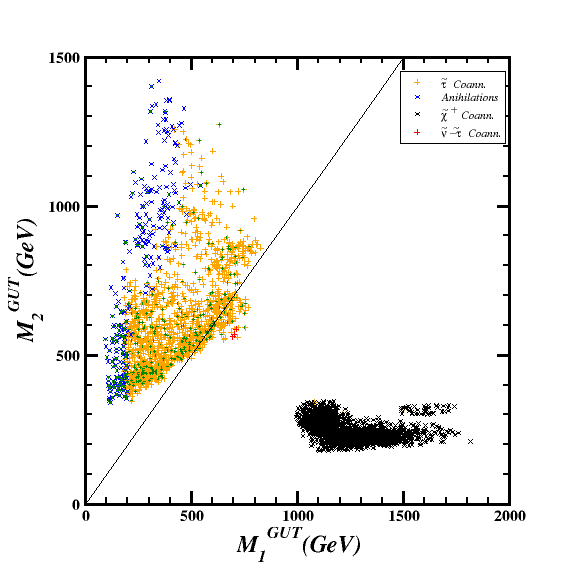}     \hspace{-.4 true cm}
        \includegraphics[scale=0.26]{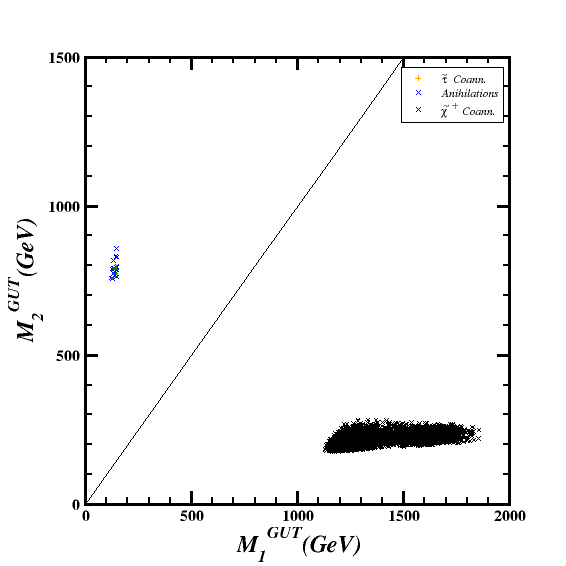}
\end{tabular}  
\caption{\footnotesize GUT values of $M_2$ vs $M_1$. Order of panels, symbols and colors are as in Figs.~\ref{fig:oh2} and \ref{fig:MLMR}. }
\label{fig:M2M1}
\end{figure}

As discussed in the previous sections, the muon $g-2$ condition imposed within $2\sigma$ band of its experimental measurements shapes the mass spectrum for the sleptons and electroweak gauginos. Fig.~\ref{fig:M2M1} shows the allowed scales for $M_{2}$ and $M_{1}$ for each scenario. As a result of the allowed cLFV processes, the parameter space is squeezed from left to right in these panels. These two soft masses play a crucial role in determining the LSP composition. As seen in all the planes, the chargino-neutralino coannihilations (black plus) mostly happen for $M_{2} < M_{1}$, where the LSP is composed mostly of Wino. Even though these solutions are not excluded by the cosmological observations, they yield a very low relic density for LSP neutralino and therefore cannot account for the observed DM abundance. If the Bino takes part in the LSP composition, the stau (and stau-sneutrino)-neutralino coannihilations can also be identified. For very light ($\lesssim 200$ GeV) $M_{1}$,  or if $M_{2}$ is relatively heavy ($\gtrsim 700$ GeV), the NLSP mass is beyond the range for suitable coannihilation scenarios and the LSP relic density is reduced through pair annihilations.  

\begin{figure}[!b]
\begin{tabular}{ccc}
\includegraphics[scale=0.26]{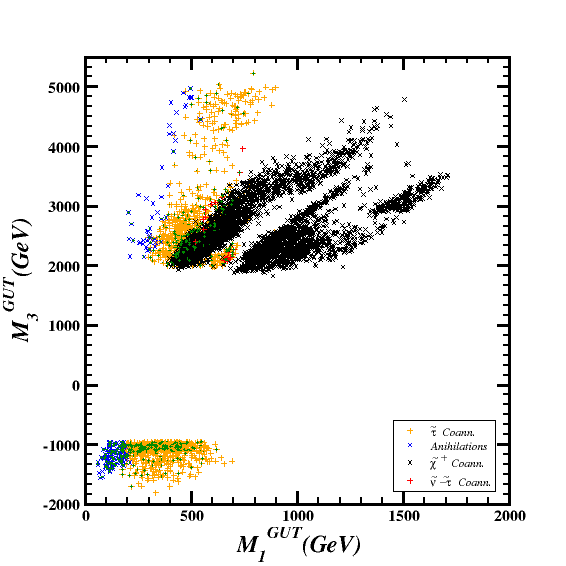}\hspace{-.4true cm}
        \includegraphics[scale=0.26]{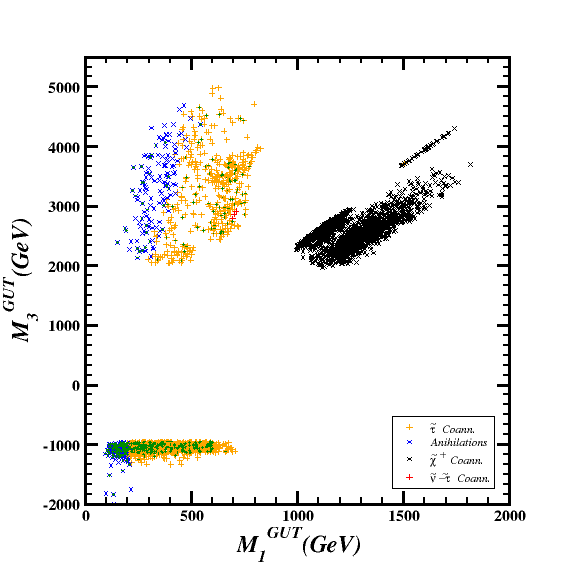}     \hspace{-.4 true cm}
        \includegraphics[scale=0.26]{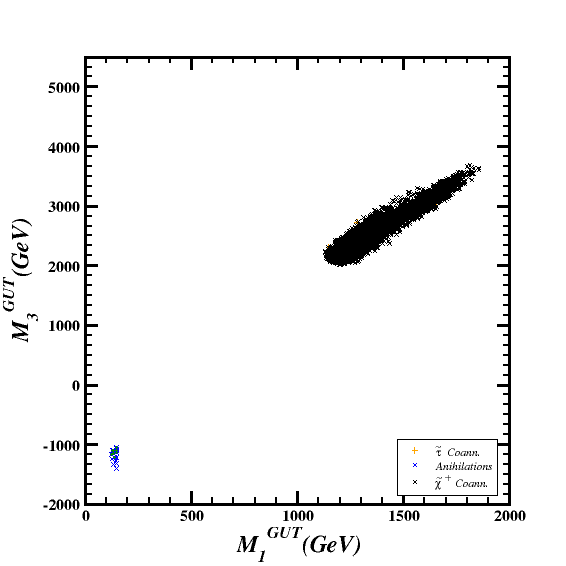}
\end{tabular}  
\caption{\footnotesize GUT values of $M_3$ vs $M_1$. Order of panels, symbols and colors are as in Figs.~\ref{fig:oh2} and \ref{fig:MLMR}.}
\label{fig:M3M1}
\end{figure}   
We have also displayed the scales for $M_{3}$ in Fig.~\ref{fig:M3M1} which summarizes the statements of suppressed flavor mixings in the quark sector due to the heavy gluinos in the regions favored by the muon $g-2$ condition. Recall that despite the light scales for $m_{L}$ and $m_{R}$, the dominant enhancement in the squark masses arises mostly from $M_{3}$ through RGEs. The large squark masses due to the heavy gluinos are also helpful in realizing the desired SM-like Higgs boson mass. To compute the Higgs boson  mass we follow the procedure in \cite{Staub:2017jnp} , as implemented in SPheno. However, in our discussion it is useful to use the approximate loop correction to the SM-like Higgs boson mass given in \cite{Carena:2012mw}, 

\begin{figure}[!t]
\centering
\begin{tabular}{ccc}
        \includegraphics[scale=0.26]{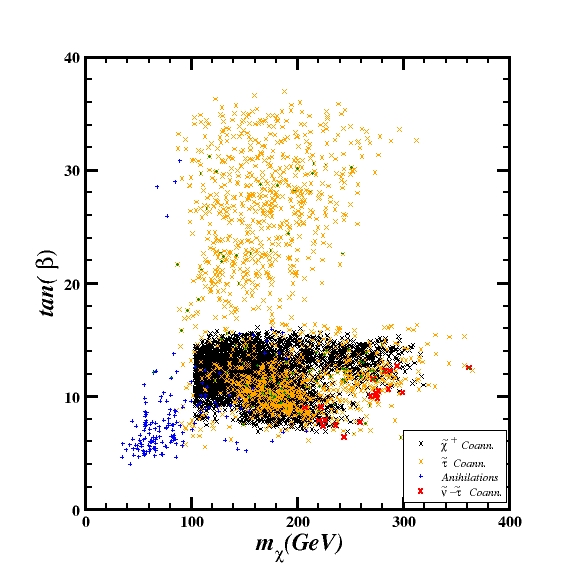}\hspace{-.4 true cm}
        \includegraphics[scale=0.26]{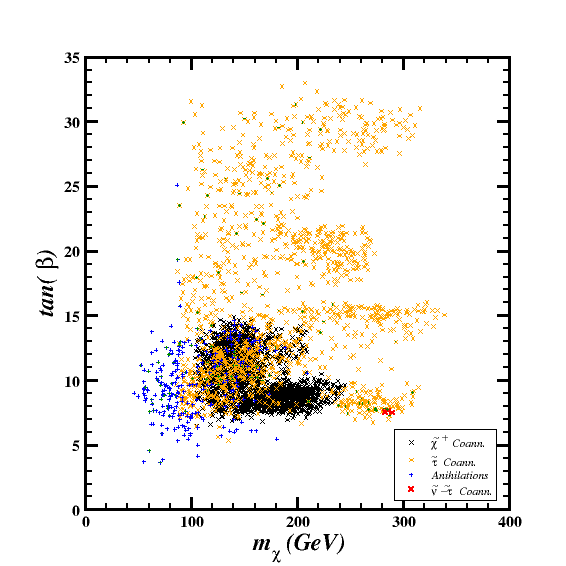}\hspace{-.4 true cm}
        \includegraphics[scale=0.26]{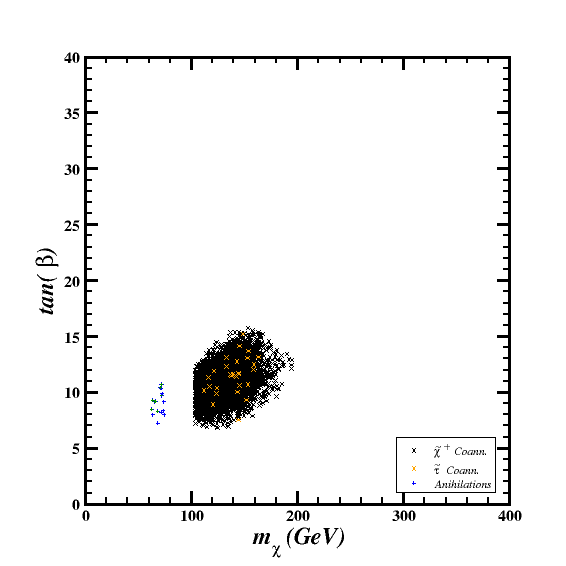}
        \end{tabular}
\caption{\footnotesize $tan\beta$ vs $m_\chi$. Order of panels, symbols and colors are as in Figs.~\ref{fig:oh2} and \ref{fig:MLMR}.}
\label{fig:tanb}
\end{figure} 

\begin{equation*}
\Delta m_{h}^{2}\simeq \dfrac{m_{t}^{4}}{16\pi^{2}v^{2}\sin^{2}\beta}\dfrac{\mu A_{t}}{M^{2}_{{\rm SUSY}}}\left[\dfrac{A_{t}^{2}}{M^{2}_{{\rm SUSY}}}-6 \right]+
\end{equation*}
\begin{equation}\hspace{1.4cm}
\dfrac{y_{b}^{4}v^{2}}{16\pi^{2}}\sin^{2}\beta\dfrac{\mu^{3}A_{b}}{M^{4}_{{\rm SUSY}}}+\dfrac{y_{\tau}^{4}v^{2}}{48\pi^{2}}\sin^{2}\beta \dfrac{\mu^{3}A_{\tau}}{m_{\tilde{\tau}}^{4}}~~,
\label{eq:higgscor}
\end{equation}
where $M_{{\rm SUSY}} \equiv \sqrt{m_{\tilde{t}_{L}}m_{\tilde{t}_{R}}}$. The first line of Eq.~\ref{eq:higgscor} represents the contributions from the stops, and the contributions depicted in the second line come from the sbottom and stau, respectively. The contribution from stops is dominant in Eq.~(\ref{eq:higgscor}), whereas the contributions from the sbottom and stau can be important for moderate and large $\tan\beta$ values. However, large $\tan\beta$ values cannot simultaneously accommodate together with the muon $g-2$ condition due to the inconsistently light SM-like Higgs boson mass (for details, see \cite{Gomez:2022qrb}). Even though $\tan\beta$ can be as high as about 40 and 35 in the models displayed in the left and middle planes of Fig. \ref{fig:tanb}, the Higgs boson mass bounds limit it to  $\tan\beta \lesssim 17$ in the models with $\epsilon = 0.2$, as shown in the right plane.

\begin{figure}[!b]
\centering
\begin{tabular}{ccc}
%\hspace{-.7 cm}
\includegraphics[scale=0.26]{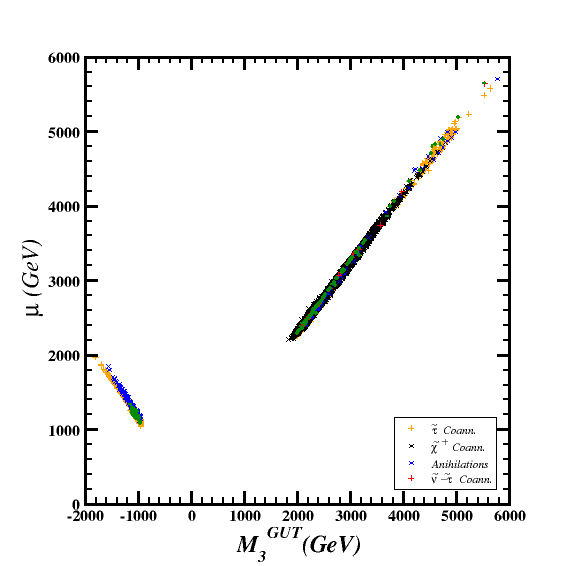}\hspace{-.4 true cm}
        \includegraphics[scale=0.26]{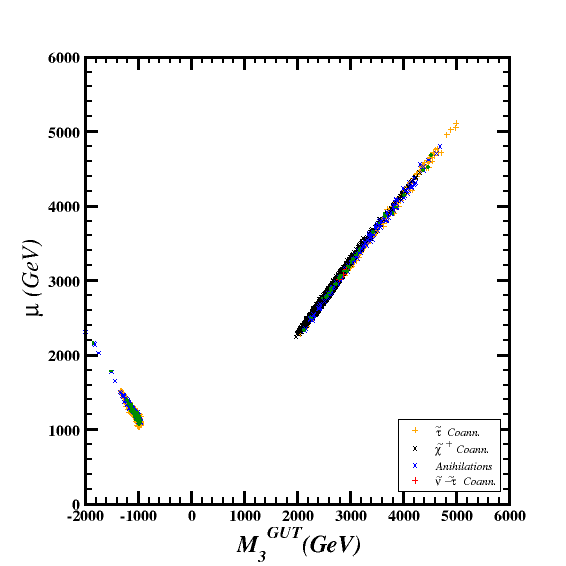}\hspace{-.4 true cm}
        \includegraphics[scale=0.26]{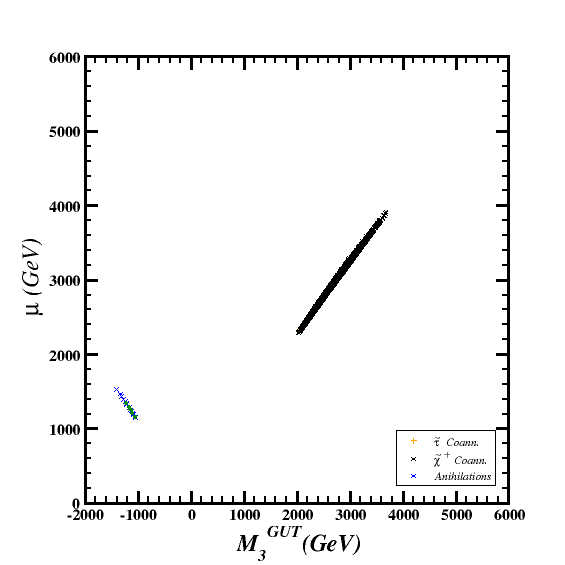}
\end{tabular}
\caption{\footnotesize 
$\mu$ vs $M_3$. Order of panels, symbols and colors are as in Figs.~\ref{fig:oh2} and \ref{fig:MLMR}.}
\label{fig:muM3}
\end{figure}   

A good prediction for Higgs boson mass favors maximal mixing of the stops \cite{Gomez:2022qrb}, but it disfavors the SUSY contribution to muon $g-2$. However, it is possible to find some values of the $\mu$-term that allow compatible predictions for both observables. For instance, in Fig.~\ref{fig:muM3} we show that some solutions can be realized for $1 \lesssim \mu \lesssim 2$ TeV, while most of the soluions yield values from 2 Tev to 6 TeV. In addition, we observe, that imposing the DM bounds requires the values of $M_3$ and $\mu$ to be strongly correlated.

\begin{figure}[t!]
\centering
%\centering
%\hline
\begin{tabular}{ccc}
\includegraphics[scale=0.26]{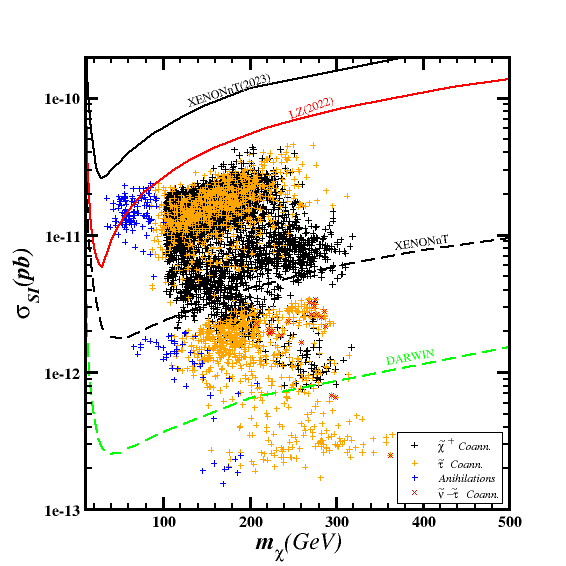}\hspace{-.4true cm}
        \includegraphics[scale=0.26]{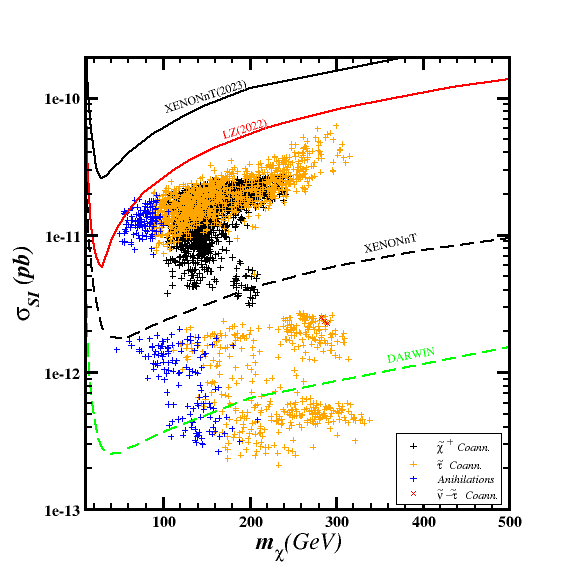}     \hspace{-.4 true cm}
        \includegraphics[scale=0.26]{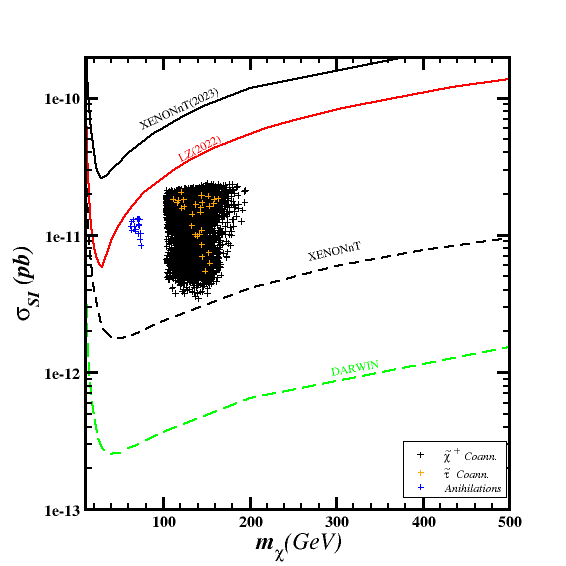}\\
\includegraphics[scale=0.26]{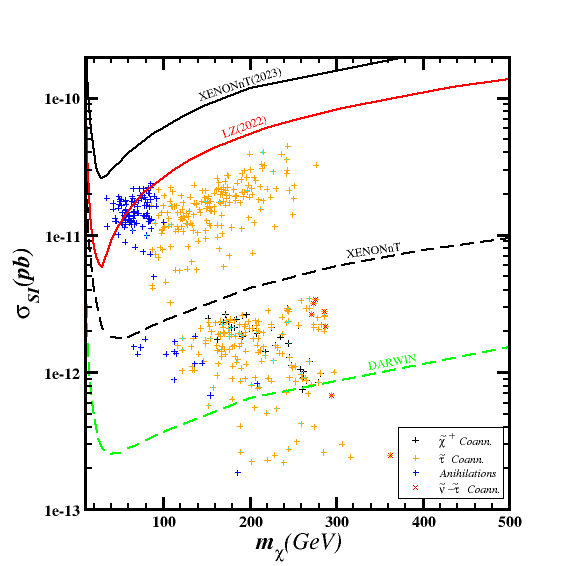}\hspace{-.4true cm}
        \includegraphics[scale=0.26]{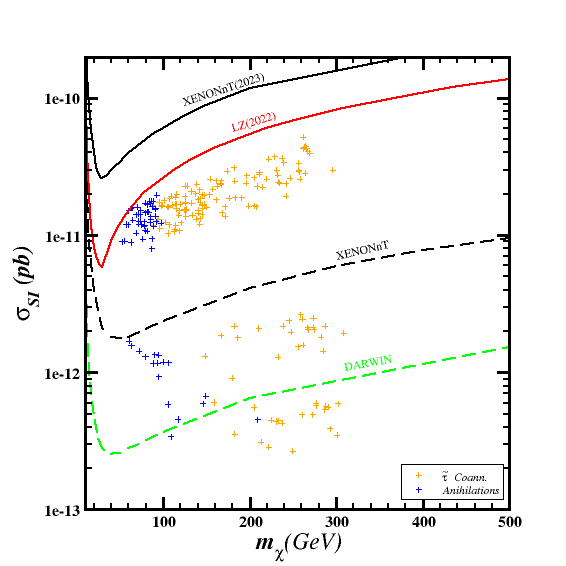}     \hspace{-.4 true cm}
        \includegraphics[scale=0.26]{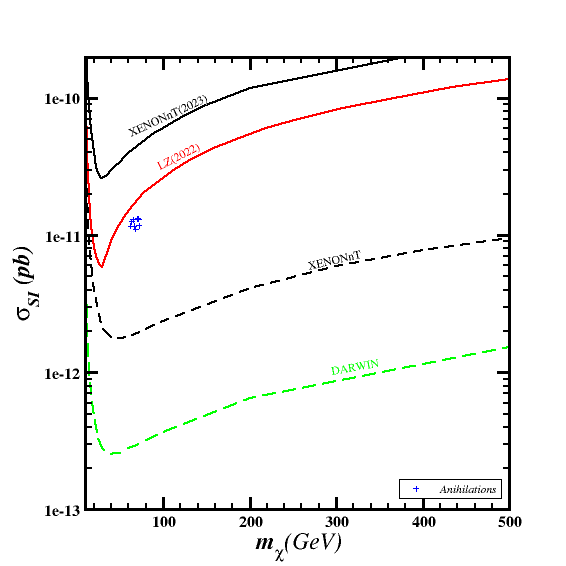}\\
\end{tabular}  
\caption{\footnotesize Spin-independent cross section in correlation with the LSP mass.  Order of panels, symbols and colors are as in Figs.~\ref{fig:oh2} and \ref{fig:MLMR}. The solid lines are bounds from XENONnT \cite{XENON:2023cxc}(black) and LZ \cite{LZ:2022lsv} (red), while dashed lines are the projected limits from XENON-nT \cite{XENON:2015gkh} (black) and DARWIN \cite{DARWIN:2016hyl}(blue). The points of the upper panels predict neutralino relic density below the upper bound in eq.~\ref{eq:Planck}, while the subset of points contained in the lower panels are also above the lower bound in eq.~\ref{eq:Planck}.}
\label{fig:sig_si}
\end{figure} 

The discussion of muon $g-2$ and SUSY mass spectra together with the DM implications should be supplemented with the possible DM detection scenarios in the direct detection experiments. In particular, the model prediction are within reach of the current sensitivities in underground experiments \cite{LZ:2022lsv,XENON:2023cxc,PandaX-4T:2021bab,DEAP:2019yzn} and the the projected limits from XENON-nT \cite{XENON:2015gkh} and DARWIN \cite{DARWIN:2016hyl}. However, the prediction for spin-dependent cross sections are lower than $10^{-7}pb$, and therefore below the current bounds from \cite{Amole_2019}. As seen from the panels of Fig.~\ref{fig:sig_si}, there are models which satisfy the DM relic density consistent with the Planck measurements. The solutions lying between the solid and the dashed curves provide a potential test for these models via the direct detection experiments of DM, and the solutions lying close to the current LZ bound are expected to be tested soon. As shown in Fig.~\ref{fig:sig_si}, most of the predictions consistent with the Planck DM bounds will be explored. Moreover, all of the models with $\epsilon=0.2$ will be tested.
\begin{figure}[!ht]
%\centering
%\hline
\begin{tabular}{ccc}
\includegraphics[scale=0.26]{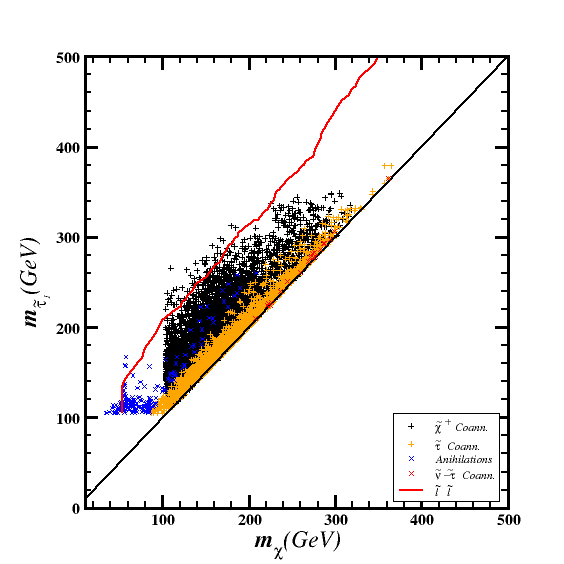}\hspace{-.4true cm}
        \includegraphics[scale=0.26]{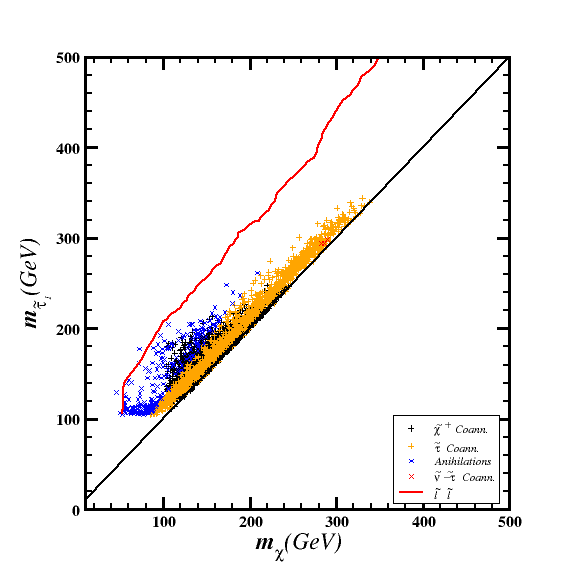}     \hspace{-.4 true cm}
        \includegraphics[scale=0.26]{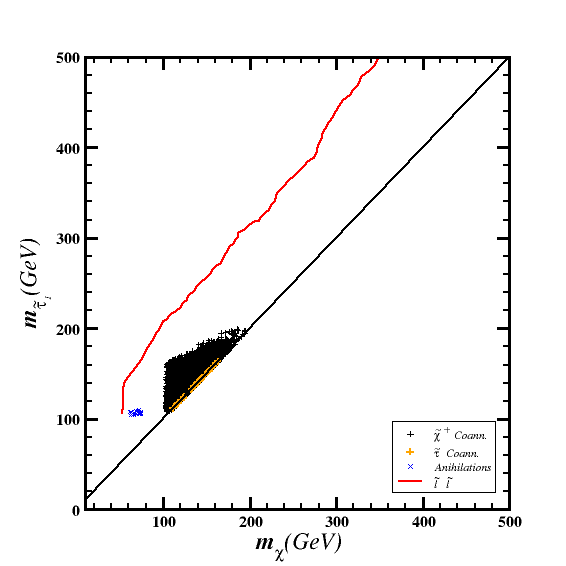}\\
\includegraphics[scale=0.26]{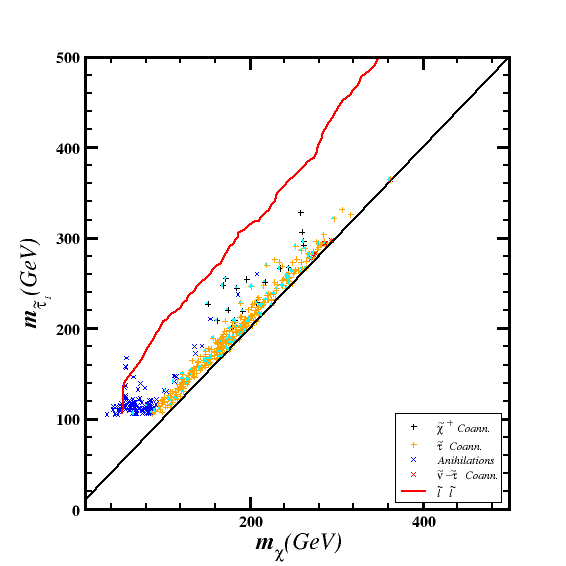}\hspace{-.4true cm}
        \includegraphics[scale=0.26]{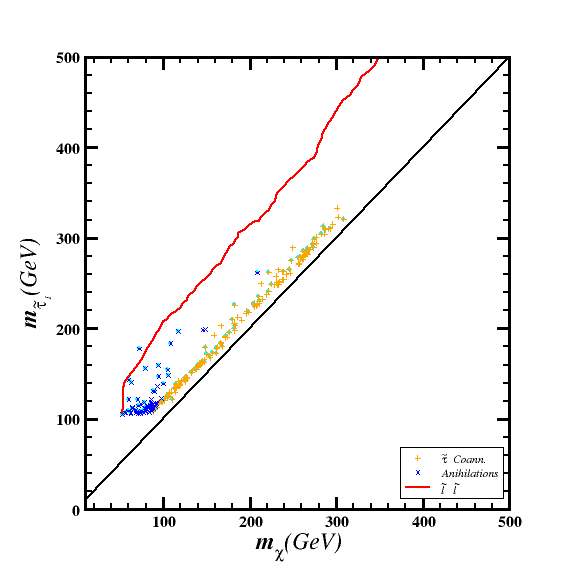}     \hspace{-.4 true cm}
        \includegraphics[scale=0.26]{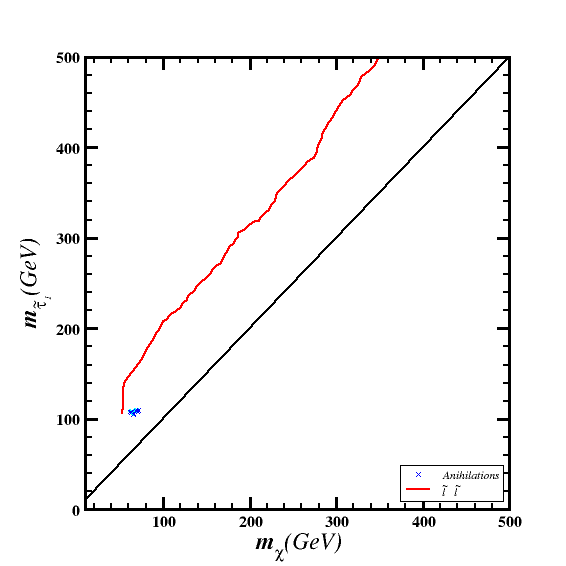}\\
\end{tabular}  
%\hline
\caption{\footnotesize $m_{\tilde{\tau}}-m_{\chi}$ plots for models with  BR($\mu\rightarrow e\gamma$) below the current bounds. Order of panels, symbols and colors are as in Figs.~\ref{fig:oh2} and \ref{fig:MLMR}. The solid lines indicate the contours extracted by the LHC searches, while the cyan marks highlight models with  BR($\mu\rightarrow e\gamma$) one order of magnitude below the current bound. The points of the upper panels predict neutralino relic density below the upper bound in eq.~\ref{eq:Planck}, while the subset of points contained in the lower panels are also above the lower bound in eq.~\ref{eq:Planck}.}
\label{fig:mstaumchi}
\end{figure}

\subsection{LFV versus LHC spectroscopy}
\label{subsec:LHC}

As seen in the previous section, the muon $g-2$ condition favors light neutralinos and charginos as well as smuons and its sneutrinos. This also implies the prediction of much lighter staus due to the larger L/R mixing in the third family. In this context, the interplay among muon $g-2$, DM and LFV constraints plays an essential role in selecting the parameter space and extracting experimental predictions. For instance, the light mass scales for stau, chargino and neutralino present an exciting opportunity for the current LHC experiments. Indeed, the light stau prediction can be tested  in the collider analyses searching for the stau-pair production in which the staus decay into the LSP neutralino together with $\tau$. Even though all of the solutions presented in Fig.~\ref{fig:mstaumchi}, fall below the bound from these analyses (red curve) due to the hadronic decays of $\tau$, the uncertainties are typically large since the desired precision in these analyses requires staus heavier than the LSP neutralino by 50 GeV or more \cite{CMS:2019hos,ATLAS:2023djh}. Thus, most of the solutions are still viable and there is considerable potential for future tests through upgrades in LHC precision. Moreover, the predictions for  BR($\mu\rightarrow e\gamma$) in many of these models are less than an order of magnitude below the current bound, and therefore observable in the coming upgrade of MEG II\cite{MEGII:2023ltw,MEGII:2023fog}.

\begin{figure}[ht!]
\centering
\begin{tabular}{ccc}
\includegraphics[scale=0.26]{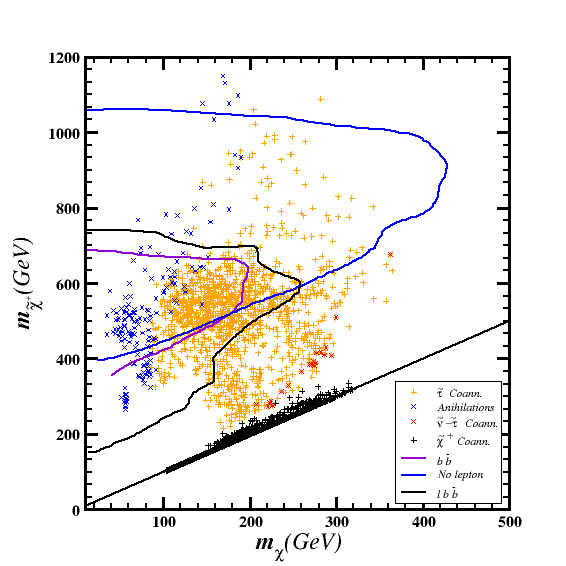}\hspace{-.4true cm}
        \includegraphics[scale=0.26]{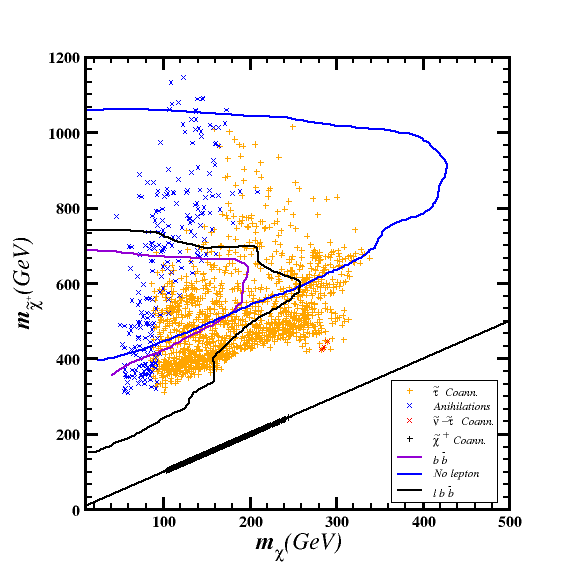}     \hspace{-.4 true cm}
        \includegraphics[scale=0.26]{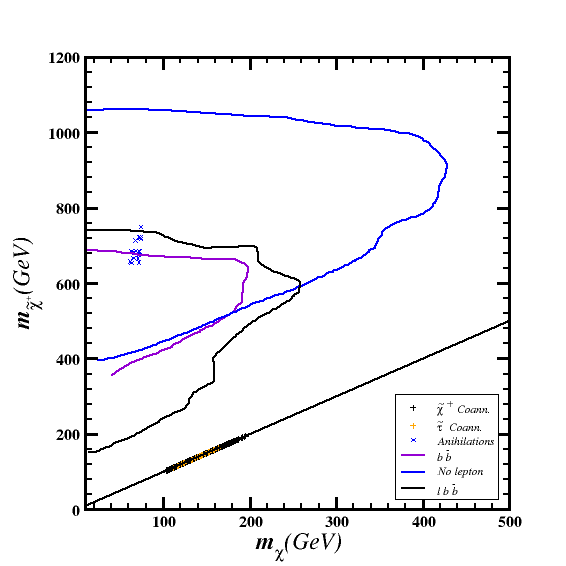}\\
\includegraphics[scale=0.26]{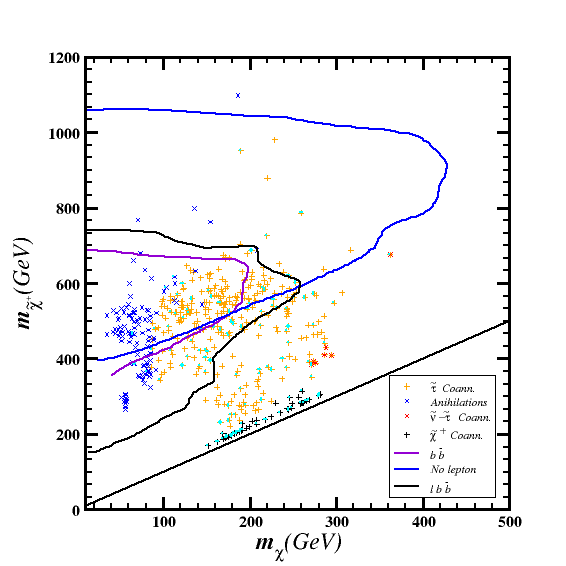}\hspace{-.4true cm}
        \includegraphics[scale=0.26]{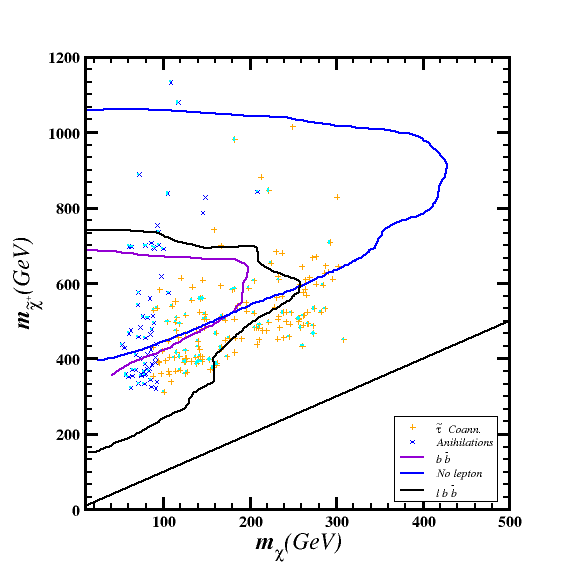}     \hspace{-.4 true cm}
        \includegraphics[scale=0.26]{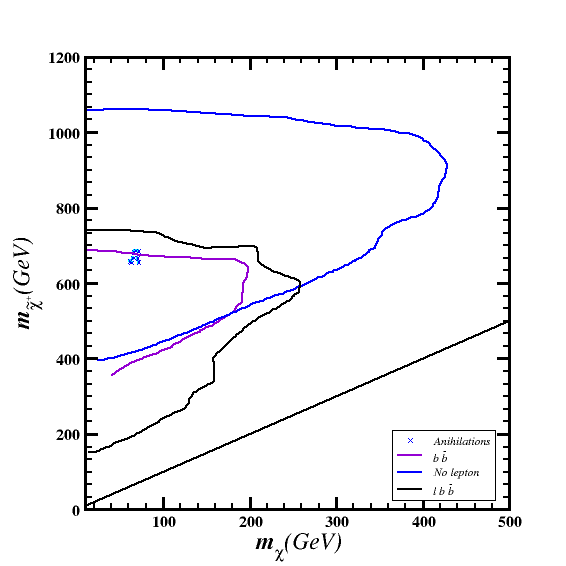}\\
\end{tabular}  
\caption{\footnotesize $m_{\tilde{\chi}^{\pm}}-m_{\chi}$ plots for models with  BR($\mu\rightarrow e\gamma$) below the current bounds. Order of panels, symbols and colors are as in Figs.~\ref{fig:oh2} and \ref{fig:MLMR}. The solid curves indicate the contours extracted by the LHC searches, while the cyan marks highlight models with  BR($\mu\rightarrow e\gamma$) one order of magnitude below the current bound. The points of the upper panels predict neutralino relic density below the upper bound in eq.~\ref{eq:Planck}, while the subset of points contained in the lower panels are also above the lower bound in eq.~\ref{eq:Planck}.}
\label{fig:mcharmchi}
\end{figure}  

A similar discussion holds for the charginos as displayed in Fig.~\ref{fig:mcharmchi} together with the bounds from the current LHC analyses. As seen here, all the models offer solutions which can be tested over a variety of different events such as those with  leptons and/or b-quarks in the the final states. Furthermore, even the most restricted models with $\epsilon = 0.2$ (right panel) provide some solutions that can be tested in $\bar{b}b$ and $l\bar{b}b$ events at the LHC. The compressed spectrum, on the other hand,  has to wait for higher precision analyses to be tested. Therefore,  the chargino-neutralino coannihilation solutions remain beyond the reach of the current collider experiments. However, the predictions for $\mu\rightarrow e \gamma$ in these models can be observable in the coming upgrade of MEG II~\cite{MEGII:2023ltw,MEGII:2023fog}.

\begin{table}[ht]
			\centering
			\setstretch{1.2}
			\scalebox{0.8}{
				\begin{tabular}{|c|c|c|c|c|}
					\hline
						
					&Ps1 & Ps2 & Ps3 & Ps4 \\ 
					\hline
					\hline
					$M_1$, $M_{2}$  &495.4, 1373.4 & 727.3, 555.7&545.3, 542.6 &649.5, 396.3 \\
					$M_3$  &4970.7& 3561.1& -970.9& 2939.7 \\
					
					\hline
					$m_{\widetilde{L}}$, $m_{\widetilde{R}}$  &275.6, 247.0 &263.7, 485.7&105.8, 246.9&369.5, 377.7  \\
					$A_{0}/m_{\widetilde{L}}$ & 0.5952& 0.8346 & 1.463& -0.6759 \\
					
					\hline
					$\tan\beta$ &15.7& 12.6 & 22.6 & 13.0 \\
					$\mu$, $m_A$  & 4994.1, 5000.7& 3743.1, 3753.5& 1103.4, 1034.3& 3219.0, 3214.1 \\
					\hline
					
					\hline
					$m_h$, $m_{H, H^{\pm}} $ &124.8, 5000 & 124.9, 3753& 124.1, 1035& 124.4, 3214 \\	
					\hline
					\hline
					$m_{\tilde{\chi}_{1}^{0}}$,$m_{\tilde{\chi}_{2}^{0}}$ &{\color{red} 186.0}, 1098.0& {\color{red}294.2}, 409.7& {\color{red}242.6}, 473.6&{\color{red} 262.3}, 283.6\\

					$m_{\tilde{\chi}_{3}^{0}}$,$m_{\tilde{\chi}_{4}^{0}}$  &5027.6, 5028.1& 3768.8, 3769.3& 1109.9, 1114.3& 3239.3, 3239.8  \\
					\hline
					$m_{\tilde{\chi}_{1}^{\pm}}$,$m_{\tilde{\chi}_{2}^{\pm}}$ & 1098.0, 5028.2& 409.9, 3769.4& 473.8, 1115.0& {\color{red}283.8}, 3240.5\\
					\hline
					$m_{\tilde{g}}$ &9773.2& 7189.0& 2141.8& 6013.3
					 \\
					\hline
					\hline
										
					$m_{\tilde{l}_1}$,$m_{\tilde{l}_2}$ & 238.2, 359.5& {\color{red}297.2}, 356.4& {\color{red}255.0}, 319.0& {\color{red}296.4}, 391.3 \\
					$m_{\tilde{l}_3}$, $m_{\tilde{l}_{4}}$ & 359.9, 814.2& 356.7, 562.1& 319.3, 387.3& 391.8, 457.4\\
					$m_{\tilde{l}_5}$, $m_{\tilde{l}_{6}}$ & 814.3, 818.4& 562.2, 572.8& 387.5, 415.0& 457.8, 498.6\\
					
					\hline
					$m_{\tilde{\nu}_{1}}$,$m_{\tilde{\nu}_{2},\tilde{\nu}_{3}}$ & 794.4, 810& {\color{red}336.7}, 348& 371.9, 379& 372.8, 384\\
					\hline
					\hline
								
					$m_{\tilde{u}_1}$,$m_{\tilde{u}_2}$ &8217.6, 8245.6& 6092.9, 6109.4& 1870.6, 1884.4& 5121.9, 5126.7 \\
					$m_{\tilde{t}_1}$,$m_{\tilde{t}_2}$ & 7129.8, 7695.7& 5297.7, 5695.0& 1593.0, 1755.1& 4412.6, 4776.5\\
					
					\hline
					$m_{\tilde{d}_1}$,$m_{\tilde{d}_2}$ &8224.2, 8246.0& 6093.4, 6112.2& 1867.4, 1886.1& 5122.6, 5129.1\\
					
					$m_{\tilde{b}_1}$,$m_{\tilde{b}_2}$ &7688.2, 8165.2& 5684.2, 6082.7& 1726.2, 1817.2& 4763.4, 5101.3 \\

										\hline
					\hline
					$\Delta a_{\mu} \times 10^{10}$ &  14.8& 15.4& 17.5& 16.3
					\\		
					\hline 			
					$\Omega h^2$ & 0.12& 0.12& 0.11& 0.09\\
					\hline

$\sigma_{SI}(pb)$ & $1.86\times 10^{-13}$ & $6.82 \times 10^{-13}$ & $3.57 \times 10^{-11}$ & $1.01 \times 10^{-12}$\\
					
$\sigma_{SD}(pb)$ & $1.12 \times 10^{-10}$ & $3.52 \times 10^{-10}$ & $5.90\times 10^{-8}$ & $6.90 \times 10^{-10}$\\
					\hline
					$BR(\mu\rightarrow e \gamma)$ & $8.7\cdot 10^{-16}$&$1.8 \cdot 10^{-14}$& $6.5\cdot 10^{-14}$& $5.7\cdot 10^{-14}$ \\
					$BR(\tau\rightarrow \mu \gamma)$& $4.8\cdot 10^{-14}$&$3.1 \cdot 10^{-13}$& $1.1\cdot 10^{-12}$& $1.2\cdot 10^{-14}$  \\
					$BR(\tau\rightarrow e \gamma)$ & $6.6\cdot 10^{-16}$&$4.2 \cdot 10^{-15}$& $1.5\cdot 10^{-14}$& $1.6\cdot 10^{-14}$
     \\
				\hline
    \hline		
			
				\end{tabular}
			}
			\caption{\footnotesize Benchmark points for models type I see-saw,  these models satisfy the same constraints of Eq.~\ref{eq:constraints}. All the points can explain the muon $g-2$ problem at $2-\sigma$ and predict neutralino relic density below the upper limit of the Planck measurements given in Eq.~\ref{eq:constraints}. We display in red the masses of the LSP and those which can coannihilate with LSP: P1 can explain relic density just with $\chi$ annihilations, and P2 corresponds to $\chi-\tilde{\tau}-\tilde{\nu}$, P3 to $\chi-\tilde{\tau}$, P4 to $\chi-\tilde{\chi}^\pm$.}
   \label{tab2}
			\end{table}
In order to make an explicit summary of the impact of LFV on signals from the MSSM models derived from $4-2-2$, we chose some representative points in the allowed solutions displayed in Figs.~\ref{fig:mstaumchi} and \ref{fig:mcharmchi}. These points can provide an indication of the magintude of the SUSY particles in models that satisfy $g-2$ due to large gluino contributions, the implication of DM in the prediction of compress SUSY spectrum and the relevance of alternative experimental signals to complement the LHC SUSY searches. Finally, we display two tables of benchmark points that are representative of our findings. The points listed in Table \ref{tab2} belong to the seesaw models, and those in Table \ref{tab3} represent the solutions realized in models with $\epsilon=0.05$ and $\epsilon = 0.2$, as indicated in the header of Table \ref{tab3}. All of the benchmark points of seesaw scenario presented in Table \ref{tab2} are compatible with the muon $g-2$ solutions within $2\sigma$, the Planck bound on the LSP neutralino relic density, and the constraint from $\mu\rightarrow e\gamma$. In addition, all these points, including P1, remain outside the contours of the current LHC bounds on the chargino and stau masses.

Similarly, the points in Table \ref{tab3} depict several coannihilation scenarios realized in the other models as summarized in the caption. Although they are compatible with the muon $g-2$ requirement and the cLFV constraints, only P1 and P5 can be consistent with the Planck measurements. The remaining solutions usually lead to LSP neutralinos with a small relic density. The chargino-neutralino coannihilation solutions, in particular, yield a relatively small LSP relic density as shown with P4 and P6. The latter is one order of magnitude smaller than P4, since it simulatenously permits stau-neutralino coannihilations and chargino-neutralino coannihilations. However, these solutions can be consistently embedded in scenarios with multi-component the DM.

		\begin{table}[ht!]
			\centering
			\setstretch{1.2}
			\scalebox{0.75}{
				\begin{tabular}{|c|c|c|c|c||c|c|}
					\hline
						
					&P1, $\epsilon=0.05$&P2,  $\epsilon=0.05$ &P3,  $\epsilon=0.05$ &P4,  $\epsilon=0.05$ &P5,  $\epsilon=0.2$ &P6,  $\epsilon=0.2$ \\ 
					\hline
					\hline
					
					$M_1$, $M_{2}$ & 375.8, 1318.6& 694.7, 561.4& 736.2, 662.1& 1552.7, 240.8& 146.6, 793.6& 1281.4, 237.2\\
					$M_3$ & 4065.4& 2788.7& 3038.3& 3052.5& -1131.3 &2172.6  \\

					\hline
	
					$m_{\widetilde{L}}$, $m_{\widetilde{R}}$  &  227.8, 185.7& 221.7, 207.4& 221.3, 233.2& 94.8, 172.4& 32.7, 96.4& 17.2, 19.4\\
				     $A_{0}/m_{\widetilde{L}}$ & 1.015& 1.31& 1.47& -0.82&0.013& -0.18  \\
					
					\hline
		
					$\tan\beta$  & 13.4& 7.5& 8.9& 13.5 & 9.7& 10.6\\
					$\mu$, $m_A$ &4172.0, 4208.3& 3021.4, 3140.3, &3243.3, 3325.8 &3312.4, 3297.3& 1240.7, 1342.5& 2448.7, 2460.0\\
						\hline
						\hline
					$m_h$, $m_{H, H^{\pm}} $  &124.9, 4208.3& 123.2, 3140.4& 123.9, 3325.9& 124.5, 3297.3&123.8, 1343.5& 123.2, 2460.0  \\	
					\hline
$m_{\tilde{\chi}_{1}^{0}}$,$m_{\tilde{\chi}_{2}^{0}}$ & {\color{red}136.9}, 1059.7 & {\color{red}282.8}, 424.6& {\color{red}300.8}, 507.9& {\color{red}144.1}, 652.5& {\color{red}72.0}, 684.9& {\color{red}156.1}, 535.9\\
					
$m_{\tilde{\chi}_{3}^{0}}$,$m_{\tilde{\chi}_{4}^{0}}$  &4200.1, 4200.7&3041.6, 3042.6& 3265.5, 3266.4& 3334.2, 3334.7&1248.9, 1254.9& 2463.9, 2464.8
 \\
					\hline
$m_{\tilde{\chi}_{1}^{\pm}}$,$m_{\tilde{\chi}_{2}^{\pm}}$ & {1059.7}, 4200.8 & {424.8}, 3043.2& {508.1}, 3266.8& {\color{red}144.3}, 3335.2&685.1, 1255.4& {\color{red}156.3}, 2465.6\\
					\hline
$m_{\tilde{g}}$ & 8086.3 & 5079.0 & 6181.5 & 6250.9& 2460.1& 4542.6\\
					\hline
					\hline

					$m_{\tilde{l}_1}$,$m_{\tilde{l}_2}$ &{ 194.11}, 280.9 & {\color{red}294.3}, 358.4 &{\color{red} 316.9}, 380.7 & 172.8, 232.0 & 110.0, 129.1& {\color{red}181.9}, 218.5 \\
					$m_{\tilde{l}_3}$, $m_{\tilde{l}_{4}}$ & 281.2,795.4 & 361.3, 363.8 & 381.2, 419.5 & 232.2, 601.8 & 129.5, 518.1& 218.6, 476.7 \\
					$m_{\tilde{l}_5}$, $m_{\tilde{l}_{6}}$ & 795.4,798.2 & 366.7,409.9 & 420.0, 457.3 & 601.8, 604.3 & 518.2, 518.7& 476.8, 482.8\\
					
					\hline
					$m_{\tilde{\nu}_{1}}$,$m_{\tilde{\nu}_{2},\tilde{\nu}_{3}}$ & 782.3, 791 & {\color{red}348.8}, 353& 406.8, 412 & 204.4,218 & 510.5, 511.8& 197.8, 203.9 \\
					\hline
					\hline
$m_{\tilde{u}_1}$,$m_{\tilde{u}_2}$& 6822.3, 6858.0& 4860.4, 4864.1& 5254.6, 5258.9& 5311.2, 5331.1& 2117.0, 2177.5& 3882.5, 3897.2\\
$m_{\tilde{t}_1}$,$m_{\tilde{t}_2}$ & 5905.6, 6405.4& 4196.3, 4553.6& 4544.6, 4922.3& 4603.6, 4956.7&1822.1, 2051.4& 3346.9, 3629.9\\
					
					\hline
$m_{\tilde{d}_1}$,$m_{\tilde{d}_2}$ & 6828.0, 6858.5& 4864.0, 4864.8& 5257.5, 5259.6& 5311.8, 5325.2& 2118.9, 2178.9& 3883.3, 3890.8 \\

$m_{\tilde{b}_1}$,$m_{\tilde{b}_2}$ & 6396.5, 6790.6& 4540.3, 4854.8& 4910.2, 5244.3& 4943.8, 5295.3& 2036.4, 2112.0& 3612.2, 3876.4\\

					\hline
					\hline
$\Delta a_{\mu} \times 10^{10}$ &  14.5 & 14.8& 14.5& 14.5 & 13.3& 13.4
 \\		
 \hline 			
 $\Omega h^2$ & 0.10 & 0.88 & 0.095 & $6.2 \times 10^{-4}$ & 0.11 & $7.4 \times 10^{-4}$\\
 \hline
$\sigma_{SI}(pb)$ & $3.22\times 10^{-13}$ & $2.52 \times 10^{-12}$ & $1.68 \times 10^{-12}$ & $6.17 \times 10^{-12}$ & $1.18 \times 10^{-11}$ & $2.07 \times 10^{-11}$ \\
					
$\sigma_{SD}(pb)$ & $2.23 \times 10^{-10}$ & $7.53\times 10^{-10}$ & $5.90 \times 10^{-8}$ & $9.9 \times 10^{-9}$ & $2.98 \times 10^{-8}$ & $3.30 \times 10^{-8}$\\
				
					\hline
					\hline
					$BR(\mu\rightarrow e \gamma)$ & $1.5 \times 10^{-13}$ & $3.9 \times 10^{-13}$ & $2.1 \times 10^{-13}$ & $2.6 \times 10^{-14}$ &$2.4 \times 10^{-13}$ & $1.1 \times 10^{-13}$  \\
					
					$BR(\tau\rightarrow \mu \gamma)$ & $1.2 \times 10^{-9}$ & $5.1 \times 10^{-11}$ & $4.3\times 10^{-11}$& $2.8 \times 10^{-12}$& $2.9\times 10^{-8}$& $6.3 \times 10^{-13}$\\
					
					$BR(\tau\rightarrow e \gamma)$ & $1.1 \times 10^{-9}$ & $1.9 \times 10^{-11}$ & $2.3 \times 10^{-11}$ & $8.5 \times 10^{-13}$ & $2.8 \times 10^{-8}$ & $4.6 \times 10^{-14}$ \\

					\hline
								\end{tabular}}
			\caption{ \footnotesize Benchmark points for models with family dependent soft terms at the GUT scale, these models satisfy the same constraints  as in Table 1. Points $P1-P4$ correspond to the models with $\epsilon=0.05$ and $P5-P6$ assume $\epsilon=0.2$. We display in red the masses of the LSP and those which can coannihilate with the LSP: P1 and P5 can explain relic density just with $\chi$ annihilations, and P2 corresponds to $\chi-\tilde{\tau}-\tilde{\nu}$, P3 to $\chi-\tilde{\tau}$, P4 and P6 to $\chi-\tilde{\chi}^\pm$ }
			\label{tab3}
		\end{table}

\section{Conclusions}
\label{sec:conc}

We have explored the predictions for LFV in a class of SUSY GUTs in which the $SO(10)$ breaks via the $4-2-2$ symmetry to the MSSM. The resulting theory does not preserve the LR symmetry among the scalar soft terms, and the universality of the gaugino masses at the GUT scale, which leads to low energy SUSY models with very interesting phenomenological predictions. For instance,  in Ref.~\cite{Gomez:2022qrb} we have shown that in this framework we are able to find models that are compatible with the measurement of the muon $g-2$ within 1-2$\sigma$, and also predict a DM relic abundance in agreement with the Planck measurements. Here, we have shown that an extension of the theory to models that can explain neutrino oscillations and/or fermion flavor leads to predictions testable in rare lepton decays, as well as a SUSY spectrum with some masses within reach of the LHC. 

In the above framework we found that a combination of SUSY masses that can explain  muon $g-2$ and with the desired DM relic density below the Planck bound can also provide LFV prdictions  in the range of current experiments. For instance, the contrast of light sfermions against heavy squarks results in measurable rates for LFV without being in conflict with the observed rates for flavor violation in the quark sector. Furthermore, even if we introduce fermion family mixings in the sfermion sector above the GUT scale,  these predict important low energy effects only in the lepton sector. The reason can be found in the large RG contribution from the gluino masses to squarks that is flavor blind and therefore washes out any generational mixings in the squark soft terms due to the growth of the diagonal terms. In contrast, the electroweak gauginos are lighter, thus allowing for generational mixings to survive in the sleptons after the RG evolution to low energies. Hence, LFV violation can be induced by non-universal soft terms associated with symmetries above the GUT scale to explain neutrino masses or fermion mass hierarchies. 

The resolution of the muon $g-2$ discrepancy within $2\sigma$ requires relatively low masses for the participating SUSY particles, which results in an upper limit of about 350 GeV on the LSP neutralino mass. This neutralino can have a composition varying from mostly Bino to mostly Wino, and depending on this composition and the mass spectrum of SUSY particles, we can find several DM scenarios. Although a Bino-like DM typically yields a large relic density, it is still possible to find scenarios where neutralino anihilations are sufficient to explain the DM relic density. However, for the majority of models, this is achieved via coannihilation with suitable SUSY particles. The $4-2-2$ symmetry establishes mass relations among SUSY particles  that favors these coannihilations,  while enhancing the SUSY contribution to  muon $g-2$. For instance, the gaugino mass relations allow neutralino-chargino coannihilations. However, these models predict a low LSP contribution to the DM relic density. For a bino-like LSP  the  LR asymmetry of the scalar masses allows sneutinos to participate in the coannihilations, thus allowing SUSY scenarios different from the ones where only staus can be the NLSP. The mass of these particles is within reach of the LHC, but characteristics of the spectrum required to satisfy muon $g-2$ work against its detection at the LHC due to their mass proximity, which makes it hard to detect their signals~\cite{CMS:2019zmn,Chakraborti:2020vjp}. In contrast, we have shown that LFV predictions for these models can be tested in the ongoing and proposed experiments. 

In summary, we presented a general seesaw model where lepton flavor is violated below the GUT scale and other cases where additional flavor symmetries may induce flavor dependent soft terms above the GUT scale. These scenarios present different signatures, that can be correlated with the DM predictions and the SUSY spectrum. Furthermore, among the benchmark points we have highlighted, LFV in $\mu$ and $\tau$ decays of experimental interest is predicted. 

\section*{Acknowledgment}
 The research of M.E.G. and C.S.U. is supported in part by the Spanish MICINN, under grant PID2022-140440NB-C22. We acknowledge Information Technologies (IT) resources at the University Of Delaware, specifically the high performance computing resources for the calculation of results presented in this paper. MEG and CSU also acknowledge the resources supporting this work in part were provided by the CEAFMC and Universidad de Huelva High Performance Computer (HPC@UHU) located in the Campus Universitario el Carmen and funded by FEDER/MINECO project UNHU-15CE-2848.

\clearpage
\bibliographystyle{JHEP}

\bibliography{LFV.bib}

\providecommand{\href}[2]{#2}\begingroup\raggedright\begin{thebibliography}{100}

\bibitem{Muong-2:2023cdq}
{\scshape Muon g-2} collaboration, \emph{{Measurement of the Positive Muon
  Anomalous Magnetic Moment to 0.20~ppm}},
  \href{https://doi.org/10.1103/PhysRevLett.131.161802}{\emph{Phys. Rev. Lett.}
  {\bfseries 131} (2023) 161802}
  [\href{https://arxiv.org/abs/2308.06230}{{\ttfamily 2308.06230}}].

\bibitem{Muong-2:2024hpx}
{\scshape Muon g-2} collaboration, \emph{{Detailed report on the measurement of
  the positive muon anomalous magnetic moment to 0.20~ppm}},
  \href{https://doi.org/10.1103/PhysRevD.110.032009}{\emph{Phys. Rev. D}
  {\bfseries 110} (2024) 032009}
  [\href{https://arxiv.org/abs/2402.15410}{{\ttfamily 2402.15410}}].

\bibitem{Muong-2:2021ojo}
{\scshape Muon g-2} collaboration, \emph{{Measurement of the Positive Muon
  Anomalous Magnetic Moment to 0.46 ppm}},
  \href{https://doi.org/10.1103/PhysRevLett.126.141801}{\emph{Phys. Rev. Lett.}
  {\bfseries 126} (2021) 141801}
  [\href{https://arxiv.org/abs/2104.03281}{{\ttfamily 2104.03281}}].

\bibitem{Muong-2:2006rrc}
{\scshape Muon g-2} collaboration, \emph{{Final Report of the Muon E821
  Anomalous Magnetic Moment Measurement at BNL}},
  \href{https://doi.org/10.1103/PhysRevD.73.072003}{\emph{Phys. Rev. D}
  {\bfseries 73} (2006) 072003}
  [\href{https://arxiv.org/abs/hep-ex/0602035}{{\ttfamily hep-ex/0602035}}].

\bibitem{ATLAS:2012yve}
{\scshape ATLAS} collaboration, \emph{{Observation of a new particle in the
  search for the Standard Model Higgs boson with the ATLAS detector at the
  LHC}}, \href{https://doi.org/10.1016/j.physletb.2012.08.020}{\emph{Phys.
  Lett. B} {\bfseries 716} (2012) 1}
  [\href{https://arxiv.org/abs/1207.7214}{{\ttfamily 1207.7214}}].

\bibitem{CMS:2013btf}
{\scshape CMS} collaboration, \emph{{Observation of a New Boson with Mass Near
  125 GeV in $pp$ Collisions at $\sqrt{s}$ = 7 and 8 TeV}},
  \href{https://doi.org/10.1007/JHEP06(2013)081}{\emph{JHEP} {\bfseries 06}
  (2013) 081} [\href{https://arxiv.org/abs/1303.4571}{{\ttfamily 1303.4571}}].

\bibitem{Chakraborti:2021dli}
M.~Chakraborti, S.~Heinemeyer and I.~Saha, \emph{{The new
  \textquotedblleft{}MUON G-2\textquotedblright{} result and supersymmetry}},
  \href{https://doi.org/10.1140/epjc/s10052-021-09900-4}{\emph{Eur. Phys. J. C}
  {\bfseries 81} (2021) 1114}
  [\href{https://arxiv.org/abs/2104.03287}{{\ttfamily 2104.03287}}].

\bibitem{Baer:2021aax}
H.~Baer, V.~Barger and H.~Serce, \emph{{Anomalous muon magnetic moment,
  supersymmetry, naturalness, LHC search limits and the landscape}},
  \href{https://doi.org/10.1016/j.physletb.2021.136480}{\emph{Phys. Lett. B}
  {\bfseries 820} (2021) 136480}
  [\href{https://arxiv.org/abs/2104.07597}{{\ttfamily 2104.07597}}].

\bibitem{Aboubrahim:2021xfi}
A.~Aboubrahim, M.~Klasen and P.~Nath, \emph{{What the Fermilab muon $g-$2
  experiment tells us about discovering supersymmetry at high luminosity and
  high energy upgrades to the LHC}},
  \href{https://doi.org/10.1103/PhysRevD.104.035039}{\emph{Phys. Rev. D}
  {\bfseries 104} (2021) 035039}
  [\href{https://arxiv.org/abs/2104.03839}{{\ttfamily 2104.03839}}].

\bibitem{Wang:2021bcx}
F.~Wang, L.~Wu, Y.~Xiao, J.M.~Yang and Y.~Zhang, \emph{{GUT-scale constrained
  SUSY in light of new muon g-2 measurement}},
  \href{https://doi.org/10.1016/j.nuclphysb.2021.115486}{\emph{Nucl. Phys. B}
  {\bfseries 970} (2021) 115486}
  [\href{https://arxiv.org/abs/2104.03262}{{\ttfamily 2104.03262}}].

\bibitem{Han:2020exx}
C.~Han, M.L.~L\'opez-Ib\'a\~nez, A.~Melis, O.~Vives, L.~Wu and J.M.~Yang,
  \emph{{LFV and (g-2) in non-universal SUSY models with light higgsinos}},
  \href{https://doi.org/10.1007/JHEP05(2020)102}{\emph{JHEP} {\bfseries 05}
  (2020) 102} [\href{https://arxiv.org/abs/2003.06187}{{\ttfamily
  2003.06187}}].

\bibitem{Altin:2017sxx}
Z.~Alt\i{}n, O.~\"Ozdal and C.S.~Un, \emph{{Muon g-2 in an alternative
  quasi-Yukawa unification with a less fine-tuned seesaw mechanism}},
  \href{https://doi.org/10.1103/PhysRevD.97.055007}{\emph{Phys. Rev. D}
  {\bfseries 97} (2018) 055007}
  [\href{https://arxiv.org/abs/1703.00229}{{\ttfamily 1703.00229}}].

\bibitem{Li:2021pnt}
Z.~Li, G.-L.~Liu, F.~Wang, J.M.~Yang and Y.~Zhang, \emph{{Gluino-SUGRA
  scenarios in light of FNAL muon g \textendash{} 2 anomaly}},
  \href{https://doi.org/10.1007/JHEP12(2021)219}{\emph{JHEP} {\bfseries 12}
  (2021) 219} [\href{https://arxiv.org/abs/2106.04466}{{\ttfamily
  2106.04466}}].

\bibitem{Ellis:2021zmg}
J.~Ellis, J.L.~Evans, N.~Nagata, D.V.~Nanopoulos and K.A.~Olive, \emph{{Flipped
  $\mathbf {g_\mu - 2}$}},
  \href{https://doi.org/10.1140/epjc/s10052-021-09829-8}{\emph{Eur. Phys. J. C}
  {\bfseries 81} (2021) 1079}
  [\href{https://arxiv.org/abs/2107.03025}{{\ttfamily 2107.03025}}].

\bibitem{Athron:2021iuf}
P.~Athron, C.~Bal\'azs, D.H.J.~Jacob, W.~Kotlarski, D.~St\"ockinger and
  H.~St\"ockinger-Kim, \emph{{New physics explanations of a$_{\mu}$ in light of
  the FNAL muon g \ensuremath{-} 2 measurement}},
  \href{https://doi.org/10.1007/JHEP09(2021)080}{\emph{JHEP} {\bfseries 09}
  (2021) 080} [\href{https://arxiv.org/abs/2104.03691}{{\ttfamily
  2104.03691}}].

\bibitem{Chakraborti:2021bmv}
M.~Chakraborti, L.~Roszkowski and S.~Trojanowski, \emph{{GUT-constrained
  supersymmetry and dark matter in light of the new $(g-2)_\mu$
  determination}}, \href{https://doi.org/10.1007/JHEP05(2021)252}{\emph{JHEP}
  {\bfseries 05} (2021) 252}
  [\href{https://arxiv.org/abs/2104.04458}{{\ttfamily 2104.04458}}].

\bibitem{Endo:2021zal}
M.~Endo, K.~Hamaguchi, S.~Iwamoto and T.~Kitahara, \emph{{Supersymmetric
  interpretation of the muon g \textendash{} 2 anomaly}},
  \href{https://doi.org/10.1007/JHEP07(2021)075}{\emph{JHEP} {\bfseries 07}
  (2021) 075} [\href{https://arxiv.org/abs/2104.03217}{{\ttfamily
  2104.03217}}].

\bibitem{Iwamoto:2021aaf}
S.~Iwamoto, T.T.~Yanagida and N.~Yokozaki, \emph{{Wino-Higgsino dark matter in
  MSSM from the g \ensuremath{-} 2 anomaly}},
  \href{https://doi.org/10.1016/j.physletb.2021.136768}{\emph{Phys. Lett. B}
  {\bfseries 823} (2021) 136768}
  [\href{https://arxiv.org/abs/2104.03223}{{\ttfamily 2104.03223}}].

\bibitem{Baum:2021qzx}
S.~Baum, M.~Carena, N.R.~Shah and C.E.M.~Wagner, \emph{{The tiny (g-2) muon
  wobble from small-$\mu$ supersymmetry}},
  \href{https://doi.org/10.1007/JHEP01(2022)025}{\emph{JHEP} {\bfseries 01}
  (2022) 025} [\href{https://arxiv.org/abs/2104.03302}{{\ttfamily
  2104.03302}}].

\bibitem{Frank:2021nkq}
M.~Frank, Y.~Hi\c{c}y\i{}lmaz, S.~Mondal, O.~\"Ozdal and C.S.~\"Un,
  \emph{{Electron and muon magnetic moments and implications for dark matter
  and model characterisation in non-universal U(1)' supersymmetric models}},
  \href{https://doi.org/10.1007/JHEP10(2021)063}{\emph{JHEP} {\bfseries 10}
  (2021) 063} [\href{https://arxiv.org/abs/2107.04116}{{\ttfamily
  2107.04116}}].

\bibitem{Heinemeyer:2021opc}
S.~Heinemeyer, E.~Kpatcha, I.n.~Lara, D.E.~L\'opez-Fogliani, C.~Mu\~noz and
  N.~Nagata, \emph{{The new $(g-2)_\mu $ result and the $\mu \nu $SSM}},
  \href{https://doi.org/10.1140/epjc/s10052-021-09601-y}{\emph{Eur. Phys. J. C}
  {\bfseries 81} (2021) 802}
  [\href{https://arxiv.org/abs/2104.03294}{{\ttfamily 2104.03294}}].

\bibitem{Akula:2013ioa}
S.~Akula and P.~Nath, \emph{{Gluino-driven radiative breaking, Higgs boson
  mass, muon g-2, and the Higgs diphoton decay in supergravity unification}},
  \href{https://doi.org/10.1103/PhysRevD.87.115022}{\emph{Phys. Rev. D}
  {\bfseries 87} (2013) 115022}
  [\href{https://arxiv.org/abs/1304.5526}{{\ttfamily 1304.5526}}].

\bibitem{Gomez:2022qrb}
M.E.~Gomez, Q.~Shafi, A.~Tiwari and C.S.~Un, \emph{{Muon $\mathbf {g-2}$,
  neutralino dark matter and stau NLSP}},
  \href{https://doi.org/10.1140/epjc/s10052-022-10507-6}{\emph{Eur. Phys. J. C}
  {\bfseries 82} (2022) 561}
  [\href{https://arxiv.org/abs/2202.06419}{{\ttfamily 2202.06419}}].

\bibitem{Ellis:2024ijt}
J.~Ellis, K.A.~Olive and V.C.~Spanos, \emph{{Non-universal SUSY models,
  $g_\mu-2$, $m_H$ and dark matter}},
  \href{https://arxiv.org/abs/2407.08679}{{\ttfamily 2407.08679}}.

\bibitem{Gomez:2010ga}
M.E.~Gomez, S.~Lola, P.~Naranjo and J.~Rodriguez-Quintero, \emph{{Suppression
  of Lepton Flavour Violation from Quantum Corrections above $M_{GUT}$}},
  \href{https://doi.org/10.1007/JHEP06(2010)053}{\emph{JHEP} {\bfseries 06}
  (2010) 053} [\href{https://arxiv.org/abs/1003.4937}{{\ttfamily 1003.4937}}].

\bibitem{Borzumati:1986qx}
F.~Borzumati and A.~Masiero, \emph{{Large Muon and electron Number Violations
  in Supergravity Theories}},
  \href{https://doi.org/10.1103/PhysRevLett.57.961}{\emph{Phys. Rev. Lett.}
  {\bfseries 57} (1986) 961}.

\bibitem{Barbieri:1995tw}
R.~Barbieri, L.J.~Hall and A.~Strumia, \emph{{Violations of lepton flavor and
  CP in supersymmetric unified theories}},
  \href{https://doi.org/10.1016/0550-3213(95)00208-A}{\emph{Nucl. Phys. B}
  {\bfseries 445} (1995) 219}
  [\href{https://arxiv.org/abs/hep-ph/9501334}{{\ttfamily hep-ph/9501334}}].

\bibitem{Goldberg:1996vd}
H.~Goldberg and M.E.~Gomez, \emph{{How Georgi-Jarlskog and SUSY SO(10) imply a
  measurable rate for mu ---\ensuremath{>} e gamma}},
  \href{https://doi.org/10.1016/S0920-5632(96)00554-3}{\emph{Nucl. Phys. B
  Proc. Suppl.} {\bfseries 52} (1997) 163}
  [\href{https://arxiv.org/abs/hep-ph/9606446}{{\ttfamily hep-ph/9606446}}].

\bibitem{Hammad:2016bng}
A.~Hammad, S.~Khalil and C.S.~Un, \emph{{Large BR$(h \to \tau \mu)$ in
  Supersymmetric Models}},
  \href{https://doi.org/10.1103/PhysRevD.95.055028}{\emph{Phys. Rev. D}
  {\bfseries 95} (2017) 055028}
  [\href{https://arxiv.org/abs/1605.07567}{{\ttfamily 1605.07567}}].

\bibitem{Abdallah:2011ew}
W.~Abdallah, A.~Awad, S.~Khalil and H.~Okada, \emph{{Muon Anomalous Magnetic
  Moment and mu -\ensuremath{>} e gamma in B-L Model with Inverse Seesaw}},
  \href{https://doi.org/10.1140/epjc/s10052-012-2108-9}{\emph{Eur. Phys. J. C}
  {\bfseries 72} (2012) 2108}
  [\href{https://arxiv.org/abs/1105.1047}{{\ttfamily 1105.1047}}].

\bibitem{Khalil:2009tm}
S.~Khalil, \emph{{Lepton flavor violation in supersymmetric B-L extension of
  the standard model}},
  \href{https://doi.org/10.1103/PhysRevD.81.035002}{\emph{Phys. Rev. D}
  {\bfseries 81} (2010) 035002}
  [\href{https://arxiv.org/abs/0907.1560}{{\ttfamily 0907.1560}}].

\bibitem{Arganda:2015naa}
E.~Arganda, M.J.~Herrero, X.~Marcano and C.~Weiland, \emph{{Enhancement of the
  lepton flavor violating Higgs boson decay rates from SUSY loops in the
  inverse seesaw model}},
  \href{https://doi.org/10.1103/PhysRevD.93.055010}{\emph{Phys. Rev. D}
  {\bfseries 93} (2016) 055010}
  [\href{https://arxiv.org/abs/1508.04623}{{\ttfamily 1508.04623}}].

\bibitem{Gomez:1998wj}
M.E.~Gomez, G.K.~Leontaris, S.~Lola and J.D.~Vergados, \emph{{U(1) textures and
  lepton flavor violation}},
  \href{https://doi.org/10.1103/PhysRevD.59.116009}{\emph{Phys. Rev. D}
  {\bfseries 59} (1999) 116009}
  [\href{https://arxiv.org/abs/hep-ph/9810291}{{\ttfamily hep-ph/9810291}}].

\bibitem{King:2004tx}
S.F.~King, I.N.R.~Peddie, G.G.~Ross, L.~Velasco-Sevilla and O.~Vives,
  \emph{{Kahler corrections and softly broken family symmetries}},
  \href{https://doi.org/10.1088/1126-6708/2005/07/049}{\emph{JHEP} {\bfseries
  07} (2005) 049} [\href{https://arxiv.org/abs/hep-ph/0407012}{{\ttfamily
  hep-ph/0407012}}].

\bibitem{Olive:2008vv}
K.A.~Olive and L.~Velasco-Sevilla, \emph{{Constraints on Supersymmetric Flavour
  Models from b ---\ensuremath{>} s gamma}},
  \href{https://doi.org/10.1088/1126-6708/2008/05/052}{\emph{JHEP} {\bfseries
  05} (2008) 052} [\href{https://arxiv.org/abs/0801.0428}{{\ttfamily
  0801.0428}}].

\bibitem{Ellis:2015dra}
S.A.R.~Ellis and G.L.~Kane, \emph{{Lepton Flavour Violation via the K\"ahler
  Potential in Compactified M-Theory}},
  \href{https://arxiv.org/abs/1505.04191}{{\ttfamily 1505.04191}}.

\bibitem{Ellis:2016qra}
J.~Ellis, K.~Olive and L.~Velasco-Sevilla, \emph{{Maximal sfermion flavour
  violation in super-GUTs}},
  \href{https://doi.org/10.1140/epjc/s10052-016-4398-9}{\emph{Eur. Phys. J. C}
  {\bfseries 76} (2016) 562}
  [\href{https://arxiv.org/abs/1605.01398}{{\ttfamily 1605.01398}}].

\bibitem{Das:2016czs}
D.~Das, M.L.~L\'opez-Ib\'a\~nez, M.J.~P\'erez and O.~Vives, \emph{{Effective
  theories of flavor and the nonuniversal MSSM}},
  \href{https://doi.org/10.1103/PhysRevD.95.035001}{\emph{Phys. Rev. D}
  {\bfseries 95} (2017) 035001}
  [\href{https://arxiv.org/abs/1607.06827}{{\ttfamily 1607.06827}}].

\bibitem{Pati:1974yy}
J.C.~Pati and A.~Salam, \emph{{Lepton Number as the Fourth Color}},
  \href{https://doi.org/10.1103/PhysRevD.10.275}{\emph{Phys. Rev. D} {\bfseries
  10} (1974) 275}.

\bibitem{Martin:2001st}
S.P.~Martin and J.D.~Wells, \emph{{Muon Anomalous Magnetic Dipole Moment in
  Supersymmetric Theories}},
  \href{https://doi.org/10.1103/PhysRevD.64.035003}{\emph{Phys. Rev. D}
  {\bfseries 64} (2001) 035003}
  [\href{https://arxiv.org/abs/hep-ph/0103067}{{\ttfamily hep-ph/0103067}}].

\bibitem{Giudice:2012pf}
G.F.~Giudice, P.~Paradisi, A.~Strumia and A.~Strumia, \emph{{Correlation
  between the Higgs Decay Rate to Two Photons and the Muon g - 2}},
  \href{https://doi.org/10.1007/JHEP10(2012)186}{\emph{JHEP} {\bfseries 10}
  (2012) 186} [\href{https://arxiv.org/abs/1207.6393}{{\ttfamily 1207.6393}}].

\bibitem{Moroi:1995yh}
T.~Moroi, \emph{{The Muon anomalous magnetic dipole moment in the minimal
  supersymmetric standard model}},
  \href{https://doi.org/10.1103/PhysRevD.53.6565}{\emph{Phys. Rev. D}
  {\bfseries 53} (1996) 6565}
  [\href{https://arxiv.org/abs/hep-ph/9512396}{{\ttfamily hep-ph/9512396}}].

\bibitem{Aoyama:2020ynm}
T.~Aoyama et~al., \emph{{The anomalous magnetic moment of the muon in the
  Standard Model}},
  \href{https://doi.org/10.1016/j.physrep.2020.07.006}{\emph{Phys. Rept.}
  {\bfseries 887} (2020) 1} [\href{https://arxiv.org/abs/2006.04822}{{\ttfamily
  2006.04822}}].

\bibitem{Lazarides:1980tg}
G.~Lazarides and Q.~Shafi, \emph{{Comments on 'Monopole Charges in Unified
  Gauge Theories'}}, {\emph{Nucl. Phys. B} {\bfseries 189} (1981) 393}.

\bibitem{Kibble:1982ae}
T.W.B.~Kibble, G.~Lazarides and Q.~Shafi, \emph{{Strings in SO(10)}},
  \href{https://doi.org/10.1016/0370-2693(82)90829-2}{\emph{Phys. Lett. B}
  {\bfseries 113} (1982) 237}.

\bibitem{Babu:1992ia}
K.S.~Babu and R.N.~Mohapatra, \emph{{Predictive neutrino spectrum in minimal
  SO(10) grand unification}},
  \href{https://doi.org/10.1103/PhysRevLett.70.2845}{\emph{Phys. Rev. Lett.}
  {\bfseries 70} (1993) 2845}
  [\href{https://arxiv.org/abs/hep-ph/9209215}{{\ttfamily hep-ph/9209215}}].

\bibitem{Anderson:1993fe}
G.~Anderson, S.~Raby, S.~Dimopoulos, L.J.~Hall and G.D.~Starkman, \emph{{A
  Systematic SO(10) operator analysis for fermion masses}},
  \href{https://doi.org/10.1103/PhysRevD.49.3660}{\emph{Phys. Rev. D}
  {\bfseries 49} (1994) 3660}
  [\href{https://arxiv.org/abs/hep-ph/9308333}{{\ttfamily hep-ph/9308333}}].

\bibitem{Drees:1986vd}
M.~Drees, \emph{{Intermediate Scale Symmetry Breaking and the Spectrum of Super
  Partners in Superstring Inspired Supergravity Models}},
  \href{https://doi.org/10.1016/0370-2693(86)90046-8}{\emph{Phys. Lett. B}
  {\bfseries 181} (1986) 279}.

\bibitem{Kawamura:1993uf}
Y.~Kawamura, H.~Murayama and M.~Yamaguchi, \emph{{Probing symmetry breaking
  pattern using sfermion masses}},
  \href{https://doi.org/10.1016/0370-2693(94)00107-3}{\emph{Phys. Lett. B}
  {\bfseries 324} (1994) 52}
  [\href{https://arxiv.org/abs/hep-ph/9402254}{{\ttfamily hep-ph/9402254}}].

\bibitem{Kolda:1995iw}
C.F.~Kolda and S.P.~Martin, \emph{{Low-energy supersymmetry with D term
  contributions to scalar masses}},
  \href{https://doi.org/10.1103/PhysRevD.53.3871}{\emph{Phys. Rev. D}
  {\bfseries 53} (1996) 3871}
  [\href{https://arxiv.org/abs/hep-ph/9503445}{{\ttfamily hep-ph/9503445}}].

\bibitem{Miller:2012vn}
D.J.~Miller, A.P.~Morais and P.N.~Pandita, \emph{{Constraining Grand
  Unification using first and second generation sfermions}},
  \href{https://doi.org/10.1103/PhysRevD.87.015007}{\emph{Phys. Rev. D}
  {\bfseries 87} (2013) 015007}
  [\href{https://arxiv.org/abs/1208.5906}{{\ttfamily 1208.5906}}].

\bibitem{Babu:2005ui}
K.S.~Babu, T.~Enkhbat and B.~Mukhopadhyaya, \emph{{Split supersymmetry from
  anomalous U(1)}},
  \href{https://doi.org/10.1016/j.nuclphysb.2005.05.006}{\emph{Nucl. Phys. B}
  {\bfseries 720} (2005) 47}
  [\href{https://arxiv.org/abs/hep-ph/0501079}{{\ttfamily hep-ph/0501079}}].

\bibitem{Milagre:2024wcg}
A.~Milagre and L.~Lavoura, \emph{{Unitarity constraints on large multiplets of
  arbitrary gauge groups}},
  \href{https://doi.org/10.1016/j.nuclphysb.2024.116542}{\emph{Nucl. Phys. B}
  {\bfseries 1004} (2024) 116542}
  [\href{https://arxiv.org/abs/2403.12914}{{\ttfamily 2403.12914}}].

\bibitem{Kibble:1982dd}
T.W.B.~Kibble, G.~Lazarides and Q.~Shafi, \emph{{Walls Bounded by Strings}},
  \href{https://doi.org/10.1103/PhysRevD.26.435}{\emph{Phys. Rev. D} {\bfseries
  26} (1982) 435}.

\bibitem{Lazarides:1985my}
G.~Lazarides and Q.~Shafi, \emph{{SUPERCONDUCTING MEMBRANES}},
  \href{https://doi.org/10.1016/0370-2693(85)90246-1}{\emph{Phys. Lett. B}
  {\bfseries 159} (1985) 261}.

\bibitem{Babu:2016bmy}
K.S.~Babu, B.~Bajc and S.~Saad, \emph{{Yukawa Sector of Minimal SO(10)
  Unification}}, \href{https://doi.org/10.1007/JHEP02(2017)136}{\emph{JHEP}
  {\bfseries 02} (2017) 136}
  [\href{https://arxiv.org/abs/1612.04329}{{\ttfamily 1612.04329}}].

\bibitem{King:1997ia}
S.F.~King and Q.~Shafi, \emph{{Minimal supersymmetric SU(4) x SU(2)-L x
  SU(2)-R}}, \href{https://doi.org/10.1016/S0370-2693(98)00058-6}{\emph{Phys.
  Lett. B} {\bfseries 422} (1998) 135}
  [\href{https://arxiv.org/abs/hep-ph/9711288}{{\ttfamily hep-ph/9711288}}].

\bibitem{Super-Kamiokande:2010orq}
{\scshape Super-Kamiokande} collaboration, \emph{{Atmospheric neutrino
  oscillation analysis with sub-leading effects in Super-Kamiokande I, II, and
  III}}, \href{https://doi.org/10.1103/PhysRevD.81.092004}{\emph{Phys. Rev. D}
  {\bfseries 81} (2010) 092004}
  [\href{https://arxiv.org/abs/1002.3471}{{\ttfamily 1002.3471}}].

\bibitem{Coriano:2014wxa}
C.~Coriano, L.~Delle~Rose and C.~Marzo, \emph{{Stability constraints of the
  scalar potential in extensions of the Standard Model with TeV scale right
  handed neutrinos}},
  \href{https://doi.org/10.1016/j.nuclphysbps.2015.06.078}{\emph{Nucl. Part.
  Phys. Proc.} {\bfseries 265-266} (2015) 311}
  [\href{https://arxiv.org/abs/1411.7168}{{\ttfamily 1411.7168}}].

\bibitem{Khalil:2010zza}
S.~Khalil and H.~Okada, \emph{{TeV scale B-L extension of the standard model}},
  \href{https://doi.org/10.1143/PTPS.180.35}{\emph{Prog. Theor. Phys. Suppl.}
  {\bfseries 180} (2010) 35}.

\bibitem{Abbas:2007ag}
M.~Abbas and S.~Khalil, \emph{{Neutrino masses, mixing and leptogenesis in TeV
  scale $B$ - L extension of the standard model}},
  \href{https://doi.org/10.1088/1126-6708/2008/04/056}{\emph{JHEP} {\bfseries
  04} (2008) 056} [\href{https://arxiv.org/abs/0707.0841}{{\ttfamily
  0707.0841}}].

\bibitem{Gomez:2020gav}
M.E.~G\'omez, Q.~Shafi and C.S.~Un, \emph{{Testing Yukawa Unification at LHC
  Run-3 and HL-LHC}},
  \href{https://doi.org/10.1007/JHEP07(2020)096}{\emph{JHEP} {\bfseries 07}
  (2020) 096} [\href{https://arxiv.org/abs/2002.07517}{{\ttfamily
  2002.07517}}].

\bibitem{Calibbi:2017uvl}
L.~Calibbi and G.~Signorelli, \emph{{Charged Lepton Flavour Violation: An
  Experimental and Theoretical Introduction}},
  \href{https://doi.org/10.1393/ncr/i2018-10144-0}{\emph{Riv. Nuovo Cim.}
  {\bfseries 41} (2018) 71} [\href{https://arxiv.org/abs/1709.00294}{{\ttfamily
  1709.00294}}].

\bibitem{Vicente:2015cka}
A.~Vicente, \emph{{Lepton flavor violation beyond the MSSM}},
  \href{https://doi.org/10.1155/2015/686572}{\emph{Adv. High Energy Phys.}
  {\bfseries 2015} (2015) 686572}
  [\href{https://arxiv.org/abs/1503.08622}{{\ttfamily 1503.08622}}].

\bibitem{Isidori:2006qy}
G.~Isidori, F.~Mescia, P.~Paradisi, C.~Smith and S.~Trine, \emph{{Exploring the
  flavour structure of the MSSM with rare K decays}},
  \href{https://doi.org/10.1088/1126-6708/2006/08/064}{\emph{JHEP} {\bfseries
  08} (2006) 064} [\href{https://arxiv.org/abs/hep-ph/0604074}{{\ttfamily
  hep-ph/0604074}}].

\bibitem{Altmannshofer:2009ne}
W.~Altmannshofer, A.J.~Buras, S.~Gori, P.~Paradisi and D.M.~Straub,
  \emph{{Anatomy and Phenomenology of FCNC and CPV Effects in SUSY Theories}},
  \href{https://doi.org/10.1016/j.nuclphysb.2009.12.019}{\emph{Nucl. Phys. B}
  {\bfseries 830} (2010) 17} [\href{https://arxiv.org/abs/0909.1333}{{\ttfamily
  0909.1333}}].

\bibitem{Crivellin:2017gks}
A.~Crivellin, G.~D'Ambrosio, T.~Kitahara and U.~Nierste, \emph{{$K\to \pi
  \nu\overline{\nu}$ in the MSSM in light of the
  $\epsilon^{\prime}_K/\epsilon_K$ anomaly}},
  \href{https://doi.org/10.1103/PhysRevD.96.015023}{\emph{Phys. Rev. D}
  {\bfseries 96} (2017) 015023}
  [\href{https://arxiv.org/abs/1703.05786}{{\ttfamily 1703.05786}}].

\bibitem{Endo:2017ums}
M.~Endo, T.~Goto, T.~Kitahara, S.~Mishima, D.~Ueda and K.~Yamamoto,
  \emph{{Gluino-mediated electroweak penguin with flavor-violating trilinear
  couplings}}, \href{https://doi.org/10.1007/JHEP04(2018)019}{\emph{JHEP}
  {\bfseries 04} (2018) 019}
  [\href{https://arxiv.org/abs/1712.04959}{{\ttfamily 1712.04959}}].

\bibitem{Stockinger:2006zn}
D.~Stockinger, \emph{{The Muon Magnetic Moment and Supersymmetry}},
  \href{https://doi.org/10.1088/0954-3899/34/2/R01}{\emph{J. Phys. G}
  {\bfseries 34} (2007) R45}
  [\href{https://arxiv.org/abs/hep-ph/0609168}{{\ttfamily hep-ph/0609168}}].

\bibitem{Cho:2011rk}
G.-C.~Cho, K.~Hagiwara, Y.~Matsumoto and D.~Nomura, \emph{{The MSSM confronts
  the precision electroweak data and the muon g-2}},
  \href{https://doi.org/10.1007/JHEP11(2011)068}{\emph{JHEP} {\bfseries 11}
  (2011) 068} [\href{https://arxiv.org/abs/1104.1769}{{\ttfamily 1104.1769}}].

\bibitem{Fargnoli:2013zia}
H.~Fargnoli, C.~Gnendiger, S.~Pa\ss{}ehr, D.~St\"ockinger and
  H.~St\"ockinger-Kim, \emph{{Two-loop corrections to the muon magnetic moment
  from fermion/sfermion loops in the MSSM: detailed results}},
  \href{https://doi.org/10.1007/JHEP02(2014)070}{\emph{JHEP} {\bfseries 02}
  (2014) 070} [\href{https://arxiv.org/abs/1311.1775}{{\ttfamily 1311.1775}}].

\bibitem{Planck:2018nkj}
{\scshape Planck} collaboration, \emph{{Planck 2018 results. I. Overview and
  the cosmological legacy of Planck}},
  \href{https://doi.org/10.1051/0004-6361/201833880}{\emph{Astron. Astrophys.}
  {\bfseries 641} (2020) A1}
  [\href{https://arxiv.org/abs/1807.06205}{{\ttfamily 1807.06205}}].

\bibitem{Carena:2012mw}
M.~Carena, S.~Gori, I.~Low, N.R.~Shah and C.E.M.~Wagner, \emph{{Vacuum
  Stability and Higgs Diphoton Decays in the MSSM}},
  \href{https://doi.org/10.1007/JHEP02(2013)114}{\emph{JHEP} {\bfseries 02}
  (2013) 114} [\href{https://arxiv.org/abs/1211.6136}{{\ttfamily 1211.6136}}].

\bibitem{Ibanez:1984vq}
L.E.~Ibanez, C.~Lopez and C.~Munoz, \emph{{The Low-Energy Supersymmetric
  Spectrum According to N=1 Supergravity Guts}},
  \href{https://doi.org/10.1016/0550-3213(85)90393-1}{\emph{Nucl. Phys. B}
  {\bfseries 256} (1985) 218}.

\bibitem{Martin:1997ns}
S.P.~Martin, \emph{{A Supersymmetry primer}},
  \href{https://doi.org/10.1142/9789812839657_0001}{\emph{Adv. Ser. Direct.
  High Energy Phys.} {\bfseries 18} (1998) 1}
  [\href{https://arxiv.org/abs/hep-ph/9709356}{{\ttfamily hep-ph/9709356}}].

\bibitem{Porod:2003um}
W.~Porod, \emph{{SPheno, a program for calculating supersymmetric spectra, SUSY
  particle decays and SUSY particle production at e+ e- colliders}},
  \href{https://doi.org/10.1016/S0010-4655(03)00222-4}{\emph{Comput. Phys.
  Commun.} {\bfseries 153} (2003) 275}
  [\href{https://arxiv.org/abs/hep-ph/0301101}{{\ttfamily hep-ph/0301101}}].

\bibitem{Porod:2011nf}
W.~Porod and F.~Staub, \emph{{SPheno 3.1: Extensions including flavour,
  CP-phases and models beyond the MSSM}},
  \href{https://doi.org/10.1016/j.cpc.2012.05.021}{\emph{Comput. Phys. Commun.}
  {\bfseries 183} (2012) 2458}
  [\href{https://arxiv.org/abs/1104.1573}{{\ttfamily 1104.1573}}].

\bibitem{Aebischer:2022vky}
J.~Aebischer, A.J.~Buras and J.~Kumar, \emph{{On the Importance of Rare Kaon
  Decays: A Snowmass 2021 White Paper}},  in \emph{{Snowmass 2021}}, 3, 2022
  [\href{https://arxiv.org/abs/2203.09524}{{\ttfamily 2203.09524}}].

\bibitem{Hisano:1995cp}
J.~Hisano, T.~Moroi, K.~Tobe and M.~Yamaguchi, \emph{{Lepton flavor violation
  via right-handed neutrino Yukawa couplings in supersymmetric standard
  model}}, \href{https://doi.org/10.1103/PhysRevD.53.2442}{\emph{Phys. Rev. D}
  {\bfseries 53} (1996) 2442}
  [\href{https://arxiv.org/abs/hep-ph/9510309}{{\ttfamily hep-ph/9510309}}].

\bibitem{Ibanez:1994ig}
L.E.~Ibanez and G.G.~Ross, \emph{{Fermion masses and mixing angles from gauge
  symmetries}}, \href{https://doi.org/10.1016/0370-2693(94)90865-6}{\emph{Phys.
  Lett. B} {\bfseries 332} (1994) 100}
  [\href{https://arxiv.org/abs/hep-ph/9403338}{{\ttfamily hep-ph/9403338}}].

\bibitem{Lola:1999un}
S.~Lola and G.G.~Ross, \emph{{Neutrino masses from U(1) symmetries and the
  Super-Kamiokande data}},
  \href{https://doi.org/10.1016/S0550-3213(99)00150-9}{\emph{Nucl. Phys. B}
  {\bfseries 553} (1999) 81}
  [\href{https://arxiv.org/abs/hep-ph/9902283}{{\ttfamily hep-ph/9902283}}].

\bibitem{Ellis:1998rj}
J.R.~Ellis, S.~Lola and G.G.~Ross, \emph{{Hierarchies of R violating
  interactions from family symmetries}},
  \href{https://doi.org/10.1016/S0550-3213(98)00423-4}{\emph{Nucl. Phys. B}
  {\bfseries 526} (1998) 115}
  [\href{https://arxiv.org/abs/hep-ph/9803308}{{\ttfamily hep-ph/9803308}}].

\bibitem{Kane:2005va}
G.L.~Kane, S.F.~King, I.N.R.~Peddie and L.~Velasco-Sevilla, \emph{{Study of
  theory and phenomenology of some classes of family symmetry and unification
  models}}, \href{https://doi.org/10.1088/1126-6708/2005/08/083}{\emph{JHEP}
  {\bfseries 08} (2005) 083}
  [\href{https://arxiv.org/abs/hep-ph/0504038}{{\ttfamily hep-ph/0504038}}].

\bibitem{King:2000ge}
S.F.~King and M.~Oliveira, \emph{{Neutrino masses and mixing angles in a
  realistic string inspired model}},
  \href{https://doi.org/10.1103/PhysRevD.63.095004}{\emph{Phys. Rev. D}
  {\bfseries 63} (2001) 095004}
  [\href{https://arxiv.org/abs/hep-ph/0009287}{{\ttfamily hep-ph/0009287}}].

\bibitem{Dent:2004dn}
T.~Dent, G.~Leontaris and J.~Rizos, \emph{{Fermion masses and proton decay in
  string-inspired SU(4) x SU(2)**2 x U(1)(X)}},
  \href{https://doi.org/10.1016/j.physletb.2004.11.032}{\emph{Phys. Lett. B}
  {\bfseries 605} (2005) 399}
  [\href{https://arxiv.org/abs/hep-ph/0407151}{{\ttfamily hep-ph/0407151}}].

\bibitem{Froggatt:1978nt}
C.D.~Froggatt and H.B.~Nielsen, \emph{{Hierarchy of Quark Masses, Cabibbo
  Angles and CP Violation}},
  \href{https://doi.org/10.1016/0550-3213(79)90316-X}{\emph{Nucl. Phys. B}
  {\bfseries 147} (1979) 277}.

\bibitem{Ellis:2020jfc}
J.~Ellis, M.E.~Gomez, S.~Lola, R.~Ruiz~de Austri and Q.~Shafi,
  \emph{{Confronting Grand Unification with Lepton Flavour Violation, Dark
  Matter and LHC Data}},
  \href{https://doi.org/10.1007/JHEP09(2020)197}{\emph{JHEP} {\bfseries 09}
  (2020) 197} [\href{https://arxiv.org/abs/2002.11057}{{\ttfamily
  2002.11057}}].

\bibitem{Casas:2001sr}
J.A.~Casas and A.~Ibarra, \emph{{Oscillating neutrinos and $\mu \to e,
  \gamma$}}, \href{https://doi.org/10.1016/S0550-3213(01)00475-8}{\emph{Nucl.
  Phys. B} {\bfseries 618} (2001) 171}
  [\href{https://arxiv.org/abs/hep-ph/0103065}{{\ttfamily hep-ph/0103065}}].

\bibitem{ParticleDataGroup:2022pth}
{\scshape Particle Data Group} collaboration, \emph{{Review of Particle
  Physics}}, \href{https://doi.org/10.1093/ptep/ptac097}{\emph{PTEP} {\bfseries
  2022} (2022) 083C01}.

\bibitem{Cannoni:2013gq}
M.~Cannoni, J.~Ellis, M.E.~Gomez and S.~Lola, \emph{{Neutrino textures and
  charged lepton flavour violation in light of $\theta_{13}$, MEG and LHC
  data}}, \href{https://doi.org/10.1103/PhysRevD.88.075005}{\emph{Phys. Rev. D}
  {\bfseries 88} (2013) 075005}
  [\href{https://arxiv.org/abs/1301.6002}{{\ttfamily 1301.6002}}].

\bibitem{Un:2016hji}
C.S.~Un and O.~Ozdal, \emph{{Mass Spectrum and Higgs Profile in BLSSM}},
  \href{https://doi.org/10.1103/PhysRevD.93.055024}{\emph{Phys. Rev. D}
  {\bfseries 93} (2016) 055024}
  [\href{https://arxiv.org/abs/1601.02494}{{\ttfamily 1601.02494}}].

\bibitem{Staub:2008uz}
F.~Staub, \emph{{SARAH}},  \href{https://arxiv.org/abs/0806.0538}{{\ttfamily
  0806.0538}}.

\bibitem{Ellwanger:1999bv}
U.~Ellwanger and C.~Hugonie, \emph{{Constraints from charge and color breaking
  minima in the (M+1)SSM}},
  \href{https://doi.org/10.1016/S0370-2693(99)00546-8}{\emph{Phys. Lett. B}
  {\bfseries 457} (1999) 299}
  [\href{https://arxiv.org/abs/hep-ph/9902401}{{\ttfamily hep-ph/9902401}}].

\bibitem{Camargo-Molina:2013qva}
J.E.~Camargo-Molina, B.~O'Leary, W.~Porod and F.~Staub,
  \emph{{$\mathbf{Vevacious}$: A Tool For Finding The Global Minima Of One-Loop
  Effective Potentials With Many Scalars}},
  \href{https://doi.org/10.1140/epjc/s10052-013-2588-2}{\emph{Eur. Phys. J. C}
  {\bfseries 73} (2013) 2588}
  [\href{https://arxiv.org/abs/1307.1477}{{\ttfamily 1307.1477}}].

\bibitem{Camargo-Molina:2013sta}
J.E.~Camargo-Molina, B.~O'Leary, W.~Porod and F.~Staub, \emph{{Stability of the
  CMSSM against sfermion VEVs}},
  \href{https://doi.org/10.1007/JHEP12(2013)103}{\emph{JHEP} {\bfseries 12}
  (2013) 103} [\href{https://arxiv.org/abs/1309.7212}{{\ttfamily 1309.7212}}].

\bibitem{Belanger:2009ti}
G.~Belanger, F.~Boudjema, A.~Pukhov and R.K.~Singh, \emph{{Constraining the
  MSSM with universal gaugino masses and implication for searches at the LHC}},
  \href{https://doi.org/10.1088/1126-6708/2009/11/026}{\emph{JHEP} {\bfseries
  11} (2009) 026} [\href{https://arxiv.org/abs/0906.5048}{{\ttfamily
  0906.5048}}].

\bibitem{Baer:2008jn}
H.~Baer, S.~Kraml, S.~Sekmen and H.~Summy, \emph{{Dark matter allowed scenarios
  for Yukawa-unified SO(10) SUSY GUTs}},
  \href{https://doi.org/10.1088/1126-6708/2008/03/056}{\emph{JHEP} {\bfseries
  03} (2008) 056} [\href{https://arxiv.org/abs/0801.1831}{{\ttfamily
  0801.1831}}].

\bibitem{Trotta:2008bp}
R.~Trotta, F.~Feroz, M.P.~Hobson, L.~Roszkowski and R.~Ruiz~de Austri,
  \emph{{The Impact of priors and observables on parameter inferences in the
  Constrained MSSM}},
  \href{https://doi.org/10.1088/1126-6708/2008/12/024}{\emph{JHEP} {\bfseries
  12} (2008) 024} [\href{https://arxiv.org/abs/0809.3792}{{\ttfamily
  0809.3792}}].

\bibitem{Staub:2017jnp}
F.~Staub and W.~Porod, \emph{{Improved predictions for intermediate and heavy
  Supersymmetry in the MSSM and beyond}},
  \href{https://doi.org/10.1140/epjc/s10052-017-4893-7}{\emph{Eur. Phys. J. C}
  {\bfseries 77} (2017) 338}
  [\href{https://arxiv.org/abs/1703.03267}{{\ttfamily 1703.03267}}].

\bibitem{Belanger:2020gnr}
G.~Belanger, A.~Mjallal and A.~Pukhov, \emph{{Recasting direct detection limits
  within micrOMEGAs and implication for non-standard Dark Matter scenarios}},
  \href{https://doi.org/10.1140/epjc/s10052-021-09012-z}{\emph{Eur. Phys. J. C}
  {\bfseries 81} (2021) 239}
  [\href{https://arxiv.org/abs/2003.08621}{{\ttfamily 2003.08621}}].

\bibitem{Planck:2015fie}
{\scshape Planck} collaboration, \emph{{Planck 2015 results. XIII. Cosmological
  parameters}},
  \href{https://doi.org/10.1051/0004-6361/201525830}{\emph{Astron. Astrophys.}
  {\bfseries 594} (2016) A13}
  [\href{https://arxiv.org/abs/1502.01589}{{\ttfamily 1502.01589}}].

\bibitem{Lahanas:1986uc}
A.B.~Lahanas and D.V.~Nanopoulos, \emph{{The Road to No Scale Supergravity}},
  \href{https://doi.org/10.1016/0370-1573(87)90034-2}{\emph{Phys. Rept.}
  {\bfseries 145} (1987) 1}.

\bibitem{Ellis:2013nka}
J.~Ellis, D.V.~Nanopoulos and K.A.~Olive, \emph{{A no-scale supergravity
  framework for sub-Planckian physics}},
  \href{https://doi.org/10.1103/PhysRevD.89.043502}{\emph{Phys. Rev. D}
  {\bfseries 89} (2014) 043502}
  [\href{https://arxiv.org/abs/1310.4770}{{\ttfamily 1310.4770}}].

\bibitem{Li:2016bww}
T.~Li, J.A.~Maxin and D.V.~Nanopoulos, \emph{{The return of the King: No-Scale
  ${\cal F}$-$SU(5)$}},
  \href{https://doi.org/10.1016/j.physletb.2016.11.022}{\emph{Phys. Lett. B}
  {\bfseries 764} (2017) 167}
  [\href{https://arxiv.org/abs/1609.06294}{{\ttfamily 1609.06294}}].

\bibitem{Ford:2019kzv}
T.~Ford, T.~Li, J.A.~Maxin and D.V.~Nanopoulos, \emph{{The heavy gluino in
  natural no-scale $\cal{F}$-$SU$(5)}},
  \href{https://doi.org/10.1016/j.physletb.2019.135038}{\emph{Phys. Lett. B}
  {\bfseries 799} (2019) 135038}
  [\href{https://arxiv.org/abs/1908.06149}{{\ttfamily 1908.06149}}].

\bibitem{Li:2021cte}
T.~Li, J.A.~Maxin and D.V.~Nanopoulos, \emph{{Spinning no-scale ${\mathcal
  {F}}$-SU(5) in the right direction}},
  \href{https://doi.org/10.1140/epjc/s10052-021-09835-w}{\emph{Eur. Phys. J. C}
  {\bfseries 81} (2021) 1059}
  [\href{https://arxiv.org/abs/2107.12843}{{\ttfamily 2107.12843}}].

\bibitem{Ellis:2020mno}
J.~Ellis, J.L.~Evans, N.~Nagata, K.A.~Olive and L.~Velasco-Sevilla,
  \emph{{Low-Energy Probes of No-Scale SU(5) Super-GUTs}},
  \href{https://doi.org/10.1140/epjc/s10052-021-08903-5}{\emph{Eur. Phys. J. C}
  {\bfseries 81} (2021) 120}
  [\href{https://arxiv.org/abs/2011.03554}{{\ttfamily 2011.03554}}].

\bibitem{MEGII:2018kmf}
{\scshape MEG II} collaboration, \emph{{The design of the MEG II experiment}},
  \href{https://doi.org/10.1140/epjc/s10052-018-5845-6}{\emph{Eur. Phys. J. C}
  {\bfseries 78} (2018) 380}
  [\href{https://arxiv.org/abs/1801.04688}{{\ttfamily 1801.04688}}].

\bibitem{MEGII:2023ltw}
{\scshape MEG II} collaboration, \emph{{A search for $\mu^+ \rightarrow e^+
  \gamma $ with the first dataset of the MEG~II experiment}},
  \href{https://doi.org/10.1140/epjc/s10052-024-12416-2}{\emph{Eur. Phys. J. C}
  {\bfseries 84} (2024) 216}
  [\href{https://arxiv.org/abs/2310.12614}{{\ttfamily 2310.12614}}].

\bibitem{MEGII:2023fog}
{\scshape MEG II} collaboration, \emph{{Operation and performance of the MEG~II
  detector}}, \href{https://doi.org/10.1140/epjc/s10052-024-12415-3}{\emph{Eur.
  Phys. J. C} {\bfseries 84} (2024) 190}
  [\href{https://arxiv.org/abs/2310.11902}{{\ttfamily 2310.11902}}].

\bibitem{Hedges:2022tnh}
{\scshape Mu2e} collaboration, \emph{{The Mu2e experiment \textemdash{}
  Searching for charged lepton flavor violation}},
  \href{https://doi.org/10.1016/j.nima.2022.167589}{\emph{Nucl. Instrum. Meth.
  A} {\bfseries 1045} (2023) 167589}
  [\href{https://arxiv.org/abs/2210.14317}{{\ttfamily 2210.14317}}].

\bibitem{COMET:2018auw}
{\scshape COMET} collaboration, \emph{{COMET Phase-I Technical Design Report}},
  \href{https://doi.org/10.1093/ptep/ptz125}{\emph{PTEP} {\bfseries 2020}
  (2020) 033C01} [\href{https://arxiv.org/abs/1812.09018}{{\ttfamily
  1812.09018}}].

\bibitem{Teshima:2019orf}
N.~Teshima, \emph{{Status of the DeeMe Experiment, an Experimental Search for
  $\mu$-$e$ Conversion at J-PARC MLF}},
  \href{https://doi.org/10.22323/1.369.0082}{\emph{PoS} {\bfseries NuFact2019}
  (2020) 082} [\href{https://arxiv.org/abs/1911.07143}{{\ttfamily
  1911.07143}}].

\bibitem{XENON:2023cxc}
{\scshape XENON} collaboration, \emph{{First Dark Matter Search with Nuclear
  Recoils from the XENONnT Experiment}},
  \href{https://doi.org/10.1103/PhysRevLett.131.041003}{\emph{Phys. Rev. Lett.}
  {\bfseries 131} (2023) 041003}
  [\href{https://arxiv.org/abs/2303.14729}{{\ttfamily 2303.14729}}].

\bibitem{LZ:2022lsv}
{\scshape LZ} collaboration, \emph{{First Dark Matter Search Results from the
  LUX-ZEPLIN (LZ) Experiment}},
  \href{https://doi.org/10.1103/PhysRevLett.131.041002}{\emph{Phys. Rev. Lett.}
  {\bfseries 131} (2023) 041002}
  [\href{https://arxiv.org/abs/2207.03764}{{\ttfamily 2207.03764}}].

\bibitem{XENON:2015gkh}
{\scshape XENON} collaboration, \emph{{Physics reach of the XENON1T dark matter
  experiment}},
  \href{https://doi.org/10.1088/1475-7516/2016/04/027}{\emph{JCAP} {\bfseries
  04} (2016) 027} [\href{https://arxiv.org/abs/1512.07501}{{\ttfamily
  1512.07501}}].

\bibitem{DARWIN:2016hyl}
{\scshape DARWIN} collaboration, \emph{{DARWIN: towards the ultimate dark
  matter detector}},
  \href{https://doi.org/10.1088/1475-7516/2016/11/017}{\emph{JCAP} {\bfseries
  11} (2016) 017} [\href{https://arxiv.org/abs/1606.07001}{{\ttfamily
  1606.07001}}].

\bibitem{PandaX-4T:2021bab}
{\scshape PandaX-4T} collaboration, \emph{{Dark Matter Search Results from the
  PandaX-4T Commissioning Run}},
  \href{https://doi.org/10.1103/PhysRevLett.127.261802}{\emph{Phys. Rev. Lett.}
  {\bfseries 127} (2021) 261802}
  [\href{https://arxiv.org/abs/2107.13438}{{\ttfamily 2107.13438}}].

\bibitem{DEAP:2019yzn}
{\scshape DEAP} collaboration, \emph{{Search for dark matter with a 231-day
  exposure of liquid argon using DEAP-3600 at SNOLAB}},
  \href{https://doi.org/10.1103/PhysRevD.100.022004}{\emph{Phys. Rev. D}
  {\bfseries 100} (2019) 022004}
  [\href{https://arxiv.org/abs/1902.04048}{{\ttfamily 1902.04048}}].

\bibitem{Amole_2019}
C.~Amole, M.~Ardid, I.~Arnquist, D.~Asner, D.~Baxter, E.~Behnke et~al.,
  \emph{Dark matter search results from the complete exposure of the pico-60
  c3f8 bubble chamber},
  \href{https://doi.org/10.1103/physrevd.100.022001}{\emph{Physical Review D}
  {\bfseries 100} (2019) }.

\bibitem{CMS:2019hos}
{\scshape CMS} collaboration, \emph{{Search for direct $\tau$ slepton pair
  production in proton-proton collisions at $\sqrt{s}=13~\mathrm{TeV}$}},
  CMS-PAS-SUS-18-006.

\bibitem{ATLAS:2023djh}
{\scshape ATLAS} collaboration, \emph{{Search for electroweak SUSY production
  in final states with two $\tau$-leptons in $\sqrt{s} = 13$ TeV $pp$
  collisions with the ATLAS detector}},  ATLAS-CONF-2023-029.

\bibitem{CMS:2019zmn}
{\scshape CMS} collaboration, \emph{{Search for Supersymmetry with a Compressed
  Mass Spectrum in Events with a Soft $\tau$ Lepton, a Highly Energetic Jet,
  and Large Missing Transverse Momentum in Proton-Proton Collisions at
  $\sqrt{s}=$ TeV}},
  \href{https://doi.org/10.1103/PhysRevLett.124.041803}{\emph{Phys. Rev. Lett.}
  {\bfseries 124} (2020) 041803}
  [\href{https://arxiv.org/abs/1910.01185}{{\ttfamily 1910.01185}}].

\bibitem{Chakraborti:2020vjp}
M.~Chakraborti, S.~Heinemeyer and I.~Saha, \emph{{Improved $(g-2)_\mu$
  Measurements and Supersymmetry}},
  \href{https://doi.org/10.1140/epjc/s10052-020-08504-8}{\emph{Eur. Phys. J. C}
  {\bfseries 80} (2020) 984}
  [\href{https://arxiv.org/abs/2006.15157}{{\ttfamily 2006.15157}}].

\end{thebibliography}\endgroup

\end{document}